\documentclass{tac}
\usepackage{latexsym}
\usepackage{url,hyperref}

\title{The Information Flow Framework:\\New Architecture}
\author{Robert E. Kent}
\address{Ontologos, 550 SW Staley Dr., Pullman, WA, USA 99163}
\eaddress{rekent@ontologos.org}
\copyrightyear{2007}
\keywords{category, metatheory, ontology, metalanguage, institution, standard}
\amsclass{18A15}

\begin{document}
\maketitle

\begin{quotation}
\noindent ``\footnotesize{\emph{Philosophy cannot become scientifically healthy without an immense technical vocabulary. We can hardly imagine our great-grandsons turning over the leaves of this dictionary without amusement over the paucity of words with which their grandsires attempted to handle metaphysics and logic. Long before that day, it will have become indispensably requisite, too, that each of these terms should be confined to a single meaning which, however broad, must be free from all vagueness. This will involve a revolution in terminology; for in its present condition a philosophical thought of any precision can seldom be expressed without lengthy explanations.}}'' 
\noindent --- Charles Sanders Peirce, Collected Papers 8:169
\end{quotation}


\begin{abstract}
The Information Flow Framework (IFF) uses institution theory
as a foundation for the semantic integration of ontologies.
It represents metalogic, and as such operates at the structural level of ontologies. 
The content, form and experience of the IFF could contribute 
to the development of a standard ontology for category theory.
The foundational aspect of the IFF helps to explain the relationship between 
the fundamental concepts of set theory and category theory. 
The development of the IFF follows two design principles: 
conceptual warrant and categorical design. 
Both are limitations of the logical expression. 
Conceptual warrant limits the content of logical expression, 
by requiring us to justify the introduction of new terminology 
(and attendant axiomatizations). 
Categorical design limits the form of logical expression 
(of all mathematical concepts and constraints) to atomic expressions: 
declarations, equations or relational expressions.
The IFF is a descriptive category metatheory. 
It is descriptive, since it follows the principle of conceptual warrant;
it is categorical, since it follows the principle of categorical design; and
it is a metatheory, since it provides a framework for all theories.
\end{abstract}


\begin{figure}
\begin{center}
\begin{tabular}[b]{l}
\setlength{\unitlength}{0.75pt}
\begin{picture}(101,100)(0,0)
\put(115,88){\makebox(0,0)[l]{\sffamily\footnotesize{metashell}}}
\put(115,47){\makebox(0,0)[l]{\sffamily\footnotesize{natural part}}}
\put(115,9){\makebox(0,0)[l]{\sffamily\footnotesize{object part}}}
\put(55,98){\makebox(0,0)[l]{\ttfamily\tiny{iff}}}
\put(60,89){\makebox(0,0)[l]{\ttfamily\tiny{type}}}
\put(65,80){\makebox(0,0)[l]{\ttfamily\tiny{meta}}}
\multiput(50,58)(0,4){3}{\circle*{1}}
\multiput(50,22)(0,4){3}{\circle*{1}}
\put(50.5,44){\makebox(0,0){\ttfamily\tiny{level $n$}}}
\put(104,96){\oval(8,8)[tr]}
\put(108,76){\line(0,1){20}}
\put(104,76){\oval(8,8)[br]}
\put(104,68){\oval(8,8)[tr]}
\put(108,20){\line(0,1){48}}
\put(104,20){\oval(8,8)[br]}
\put(104,11.5){\oval(8,8)[tr]}
\put(108,4.5){\line(0,1){7}}
\put(104,4.5){\oval(8,8)[br]}
\multiput(8.6,8)(8.26,0){11}{\line(0,1){8}}
\put(4,8){\line(1,0){92}}
\multiput(5,0)(8.2,0){12}{\line(0,1){8}}
\thicklines
\put(0,0){\line(1,2){50}}
\put(100,0){\line(-1,2){50}}
\multiput(47.2,94)(2.1,0){3}{\rule{1pt}{.4pt}}
\multiput(42.1,84)(2.4,0){7}{\rule{1pt}{.4pt}}
\put(36,72){\line(1,0){28}}
\multiput(26,51)(2.9,0){17}{\rule{1pt}{.4pt}}
\multiput(18.4,36)(2.95,0){22}{\rule{1pt}{.4pt}}
\put(8,16){\line(1,0){84}}
\put(0,0){\line(1,0){100}}
\end{picture}
\end{tabular}
\end{center}
\caption{IFF Architecture (iconic)}
\label{architecture:iconic}
\end{figure}
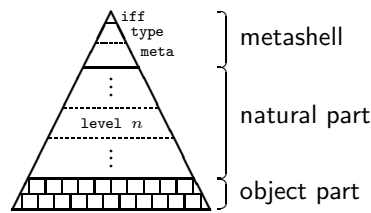

\section{Introduction}\label{sec-introduction}

The Information Flow Framework (IFF)
is a descriptive category metatheory \cite{kent:04} currently under development 
that provides an important practical application of category theory \cite{maclane:71}
to knowledge representation, knowledge maintenance and the semantic web (\cite{kent:03},\cite{kent:iswc03}).
The categorical approach of the IFF provides a principled framework for 
the modular design and semantic integration of object-level ontologies. 
The IFF forms the structural aspect of the IEEE P1600.1 Standard Upper Ontology (SUO) project\footnote{The main IFF webpage is located at \url{http://suo.ieee.org/IFF/}.}. 
It is an experiment in foundations,
which follows a bottom-up approach to logical description. 
A preliminary description of the IFF was presented at 
the International Category Theory Conference in Vancouver in 2004 \cite{kent:04}.
This paper discusses a new, modular, more mature architecture
(the transition from the preliminary description to this more mature architecture is discussed in subsubsection~\ref{subsubsec-transition}).


\subsection{Background}\label{subsec-background}

Before discussing the architecture,
we discuss:
roles in category theory,
ontologies with examples,
development, design principles and current state.

\subsubsection{Roles}\label{subsubsec-roles}

\begin{figure}
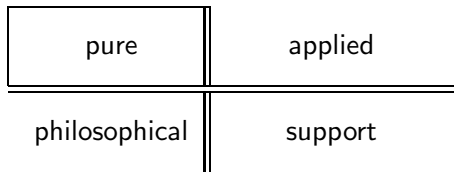

\begin{center}
\rule{30pt}{0pt}\begin{tabular}{|c||c}
\multicolumn{2}{c}{} \\ \cline{1-1}
\rule[-12pt]{0pt}{30pt}
{\sffamily\small pure}         & \makebox[80pt]{\sffamily\small applied} \\ \hline\hline 
\multicolumn{1}{c||}{\rule[-12pt]{0pt}{30pt}
\sffamily\small philosophical} & {\sffamily\small support}
\end{tabular}
\end{center}
\caption{Roles in Category Theory}
\label{roles}
\end{figure}

All category theory activities involve several roles (Figure~\ref{roles}):
pure, applied, philosophical and support.
The pure role is to develop category theory.
This is the central role. 
An example of this role was played by
the International Category Theory Conference at White Point in 2006.
The applied role uses category theory in various applications such as
mathematics, programming languages, concurrency and knowledge engineering.
The philosophical role tries to explain and/or justify category theory.
This may be based on various historical and social forces, or its position in reality.
At any particular time, an activity may emphasize a certain role.
The support role is for the implementation of category theory.
This will aid the working category theorist.
Support might involve the development of suitable ontologies, logical code, grammars and programming code.
The IFF development has involved several of these roles.
Initially,
its goal was the application of category theory to knowledge engineering.
More recently,
the IFF has reverse this approach,
and now seeks to support category theory by applying some of the tools and techniques of knowledge engineering.
Since this recent supporting role involves foundations,
the IFF is implicitly involved to a certain extent in a philosophical role.

\subsubsection{Ontologies}\label{ontologies}

The term `ontology' was first coined in the 17th century.
It comes from two Greek words, 
$o\nu\tau{o}\varsigma$ meaning \emph{of being} 
and 
$\lambda{o}\gamma \iota \alpha$ meaning \emph{science}, \emph{study} or \emph{theory}.
The ancient Greek philosopher Aristotle defined ontology to be
``the science of being qua (in the capacity of) being''; 
hence, ontology is the science of being inasmuch as it is being, 
or the study of beings insofar as they exist.
The dictionary (Merriam-Webster) defines ontology to be
{1} : 
a branch of metaphysics concerned with the nature and relations of being
{2} : 
a particular theory about the nature of being or the kinds of existents.
The Encyclopedia (Brittanica) describes ontology to be
the theory or study of being as such; 
i.e., of the basic characteristics of all reality. 
Ontology is synonymous with metaphysics or ``first philosophy'' 
as defined by Aristotle. 
Artificial Intelligence and Knowledge Engineering define an ontology\footnote{Ontologies can be thought of as taxonomies, logical theories or knowledge-bases.} 
to be a formal, explicit specification of a shared conceptualization. 
It is \emph{an abstract model of some phenomena in the world} (semantic conceptualization), explicitly represented as \emph{concepts, relationships and constraints} (logic-oriented), which is \emph{machine-readable} (formal and explicit) and incorporates \emph{the consensual knowledge of some community} (shared and relative).

\subsubsection{Examples}\label{examples}

Figure~\ref{road:maps} illustrates an ontology of roadmaps.
In the area of biology,
the Gene Ontology (GO)\footnote{\,located at 
\url{http://www.geneontology.org/}}
is an actual example of a functioning ontology,
representing 
concepts such as gene, protein and metabolic pathway,
and predicates such as being a regulator gene.
A possible ontology for the category theory community 
might represent concepts such as category or adjunction,
predicates such as small-complete, 
functions such as the object/morphism set function of a category, 
and relations such as the subcategory order or the composable relation for functors. 

{\footnotesize \begin{figure}
\begin{center}
\begin{tabular}[t]{@{\hspace{45pt}}c@{\hspace{-5pt}}c}

\begin{tabular}[t]{l}
{\tiny \begin{tabular}[t]{p{1.7in}}
\begin{description}
\item [Concepts = Types = Entities] \mbox{ }
\begin{tabular}[t]{l}
\\
$\bullet$ highway = road \\
$\bullet$ geographical-feature \\
\begin{tabular}[t]{l}
$-$ location = point \\
\begin{tabular}[t]{l}
$\ast$ exit \\
$\ast$ interchange \\
$\ast$ town \\
$\ast$ rest-area
\end{tabular} \\
$-$ line = linear-feature \\
\begin{tabular}[t]{l}
$\ast$ creek \\
$\ast$ river \\
$\ast$ railroad
\end{tabular} \\
$-$ area \\
\begin{tabular}[t]{l}
$\ast$ lake \\
$\ast$ mountain \\
$\ast$ city \\
$\ast$ county \\
$\ast$ state = province \\
$\ast$ country
\end{tabular} \\
\end{tabular} \\
$\bullet$ territorial-division \\
\begin{tabular}[t]{l}
$-$ county \\
$-$ state \\
$-$ country
\end{tabular} \\
$\bullet$ urban-area \\
\begin{tabular}[t]{l}
$-$ town \\
$-$ city
\end{tabular} \\
\end{tabular}
\end{description}
\end{tabular}}
\end{tabular}

&

\begin{tabular}[t]{l}
{\tiny \begin{tabular}[t]{p{2.7in}}
\begin{description}
\item [Predicates = Parts] \mbox{ } 
\newline\newline
{\tiny $\begin{array}[t]{r@{\hspace{5pt}:\hspace{5pt}}l}
 \mbox{principal} & \mbox{highway} \rule{0pt}{6pt} \\
 \mbox{toll-road} & \mbox{highway} \rule{0pt}{6pt} \\
   \mbox{freeway} & \mbox{highway} \rule{0pt}{6pt} \\
    \mbox{scenic} & \mbox{highway} \rule{0pt}{6pt} \\
\mbox{is-capital} & \mbox{urban-area} \rule{0pt}{6pt} 
\end{array}$}
\newline
\item [Functions = Maps] \mbox{ } 
\newline\newline
{\tiny $\begin{array}[t]{r@{\hspace{5pt}:\hspace{5pt}}l}
  \mbox{name} (\mbox{number}) & \mbox{highway} \rightarrow \mbox{name-tag} {\times} \mbox{number} \rule{0pt}{6pt} \\
\mbox{number-of-lanes} & \mbox{highway} \rightarrow \mbox{number} \rule{0pt}{6pt} \\
       \mbox{distance} & \mbox{point} {\times} \mbox{point} \rightarrow \mbox{number} \rule{0pt}{6pt} \\
     \mbox{facility} & \mbox{rest-area} \rightarrow \mbox{facility-tag} \rule{0pt}{6pt} \\
   \mbox{intersection} & \mathsf{ext}(\mbox{crosses}) \rightarrow \mbox{point} \rule{0pt}{6pt} \\
  \mbox{exit-location} & \mbox{exit} \rightarrow \mbox{highway} {\times} \mbox{number} \rule{0pt}{6pt} \\ 
        \mbox{lies-in} & \mbox{county} \rightarrow \mbox{state} \rule[-4pt]{0pt}{8pt} \\ \cline{1-1}
\multicolumn{2}{l}{\mbox{name-tag} = \{\mbox{\ttfamily interstate}, \mbox{\ttfamily state}, \mbox{\ttfamily county}\}} \rule{0pt}{11pt} \\
\multicolumn{2}{l}{\mbox{facility-tag} = \{\mbox{\ttfamily full},\mbox{\ttfamily partial},\mbox{\ttfamily none}\}} \rule{0pt}{8pt} 
\end{array}$}
\newline
\item [Relations] \mbox{ } 
\newline\newline
{\tiny $\begin{array}[t]{r@{\hspace{5pt}:\hspace{5pt}}l}
     \mbox{crosses} & \mbox{line} \rightharpoondown \mbox{line} \rule{0pt}{6pt} \\
   \mbox{traverses} & \mbox{highway} \rightharpoondown \mbox{territorial-division} \rule{0pt}{6pt} \\
\mbox{goes-through} & \mbox{road} \rightharpoondown \mbox{urban-area} \rule{0pt}{6pt}
\end{array}$}
\end{description}
\end{tabular}}
\end{tabular}

\\

\multicolumn{2}{c}{
\begin{tabular}[t]{l}
{\tiny \begin{tabular}[t]{p{3.5in}}
\begin{description}
\item [Axioms] \mbox{ } 
\newline\newline
{\tiny $\begin{array}[t]{l}
\forall_{(\scriptscriptstyle x,y \in \mbox{linear-feature})} 
\left( \mbox{crosses}(x,y) 
       \Rightarrow \mbox{crosses}(y,x) \right) \rule{0pt}{6pt}                  \\
\forall_{(h \in \mbox{highway}, c \in \mbox{county}, s \in \mbox{state})} 
\left( (\mbox{traverses}(h,c) \;\&\; \mbox{lies-in}(c,s)) 
       \Rightarrow \mbox{traverses}(h,s) \right) \rule{0pt}{6pt}                \\
\forall_{(x,y,z \in \mbox{location})} 
\left( \mbox{distance}(x,z) 
       \leq \mbox{distance}(x,y) + \mbox{distance}(y,z) \right) \rule{0pt}{6pt} 
\end{array}$}
\end{description}
\end{tabular}}
\end{tabular}}

\end{tabular}
\end{center}
\caption{An Ontology of Roadmaps}
\label{road:maps}
\end{figure}}

\subsection{The IFF}\label{subsec-iff}

The IFF originated from a desire to 
use category theory for the representation and semantic integration of ontologies.

\subsubsection{Development}\label{development}

The IFF is being develop under the auspices of 
the IEEE Standard Upper Ontology (SUO) project \cite{suo:homepage}.
It was the first of several approved SUO resolutions.
There was always a close connection \cite{kent:dagstuhl}
between the goals of the IFF and the theory of institutions \cite{goguen:burstall:92}. 
There was also a connection to foundations, 
since from the category-theoretic perspective,
a strong requirement of the IFF formalism 
was the complete incorporation of various structures 
in a large category $\mathcal{C}$, 
such as the pullback square that defines the source of the composition map
\begin{center}
\begin{tabular}{c}
\setlength{\unitlength}{0.5pt}
\begin{picture}(160,110)(0,-15)
\put(-25,70){\makebox(50,20){\footnotesize $\mathsf{mor}(\mathcal{C}) {\times}_{\mathsf{obj}(\mathcal{C})} \mathsf{mor}(\mathcal{C})$}}
\put(-25,-10){\makebox(50,20){\footnotesize $\mathsf{mor}(\mathcal{C})$}}
\put(135,70){\makebox(50,20){\footnotesize $\mathsf{mor}(\mathcal{C})$}}
\put(135,-10){\makebox(50,20){\footnotesize $\mathsf{obj}(\mathcal{C})$}}
\put(-40,30){\makebox(50,20){\scriptsize $\pi_0^{\mathcal{C}}$}}
\put(79,86){\makebox(50,20){\scriptsize $\pi_1^{\mathcal{C}}$}}
\put(156,30){\makebox(50,20){\scriptsize $\partial_0^{\mathcal{C}}$}}
\put(55,-28){\makebox(50,20){\scriptsize $\partial_1^{\mathcal{C}}$}}
\put(85,80){\vector(1,0){40}}
\put(30,0){\vector(1,0){100}}
\put(0,60){\vector(0,-1){40}}
\put(160,60){\vector(0,-1){40}}
\thinlines
\put(120,37){\line(0,-1){20}}
\put(120,37){\line(1,0){20}}
\end{picture}
\end{tabular}
\end{center}

\subsubsection{Design Principles}\label{design:principles}

During the IFF development,
two design principles have emerged as important:
conceptual warrant and categorical design.
\begin{itemize}
\item 
\noindent {Conceptual Warrant:} [content]\footnote{A non-starter during the IFF development was a topos axiomatization.
This received objections from the SUO working group, 
in part due to its lack of support by motivating examples.
Rejection of the topos axiomatization prompted the idea of conceptual warrant.}
\emph{IFF terminology requires conceptual warrant.}
Warrant means evidence for or token of authorization. 
Conceptual warrant is an adaptation of the librarianship notion of literary warrant.
According to the Library of Congress (LOC), 
its collections serve as the literary warrant 
(i.e., the literature on which the controlled vocabulary is based) 
for the LOC subject headings system.
Likewise for the IFF,
its lower or more peripheral axiomatized concepts
serve as the conceptual warrant 
for the higher level IFF terminology;
hence,
any term should reference a concept needed 
in a lower metalevel or more peripheral axiomatization.

\begin{center}
{\small \begin{tabular}{r|c|c|} \cline{2-3}
LOC & subject headings & collections    \\ \cline{2-3}
IFF & higher terms     & lower concepts \\ \cline{2-3}
\end{tabular}}
\end{center}
\item 
\noindent {Categorical Design:} [form]
\emph{IFF module design should follow good category-theoretic intuitions.}
Axiomatizations should complete any implicit ideas.
For example,
any implicit adjunctions should be formalized explicitly.
Any current axiomatization may only be partially completed.
Axiomatizations should be atomic.
Thus,
axiomatizations should be in the form of declarations, equations and relational expressions.
No axioms should use explicit logical notation:
no variables, quantifications or logical connectives should be used.
Although the metashell axiomatization uses first order expression,
the natural part axiomatization is atomic.
\end{itemize}

\subsubsection{Current State}\label{current:state}

\begin{figure}
\begin{center}
\begin{tabular}{@{\hspace{80pt}}c@{\hspace{30pt}}c}
\setlength{\unitlength}{0.4pt}
\begin{picture}(200,0)(0,-20)
\put(175,10){\begin{picture}(0,25)(0,0)
\thicklines
\put(0,25){\circle*{5}}
\put(0,25){\vector(0,-1){25}}
\end{picture}}
\put(-120,-20){\makebox(50,0)[r]{\footnotesize\begin{tabular}[t]{c}{expression}\end{tabular}}}
\put(-25,-60){\makebox(50,0){\tiny\begin{tabular}[t]{c}natural\\language\end{tabular}}}
\put(75,-60){\makebox(50,0){\tiny\begin{tabular}[t]{c}\mbox{ }\\first-order\end{tabular}}}
\put(175,-60){\makebox(50,0){\tiny\begin{tabular}[t]{c}\mbox{ }\\atomic\end{tabular}}}
\put(0,0){\begin{picture}(200,0)(0,0)
\put(-50,-30){\makebox(50,20)[r]{\scriptsize$0\,$}}
\put(52,-30){\makebox(50,20)[r]{\scriptsize$1\,$}}
\put(150,-30){\makebox(50,20)[r]{\scriptsize$2\,$}}
\multiput(0,0)(100,0){2}{\begin{picture}(100,0)(0,0)
\put(0,0){\line(1,0){100}}
\put(0,0){\line(0,-1){30}}
\put(25,0){\line(0,-1){18}}
\put(50,0){\line(0,-1){24}}
\put(75,0){\line(0,-1){18}}
\put(100,0){\line(0,-1){30}}
\end{picture}}
\end{picture}}
\end{picture}
&
{\scriptsize \begin{tabular}{p{200pt}}
\\
\begin{itemize}
\item 
Axioms in the metashell are in first order form;
most axioms in the natural part are in atomic form.
\end{itemize} 
\end{tabular}}
\\
\setlength{\unitlength}{0.4pt}
\begin{picture}(200,0)(0,-20)
\put(50,10){\begin{picture}(0,25)(0,0)
\thicklines
\put(0,25){\circle*{5}}
\put(0,25){\vector(0,-1){25}}
\end{picture}}
\put(-120,-20){\makebox(50,0)[r]{\footnotesize\begin{tabular}[t]{c}{category}\end{tabular}}}
\put(-25,-60){\makebox(50,0){\tiny\begin{tabular}[t]{c}finitely\\complete\end{tabular}}}
\put(75,-60){\makebox(50,0){\tiny\begin{tabular}[t]{c}cartesian\\closed\end{tabular}}}
\put(175,-60){\makebox(50,0){\tiny\begin{tabular}[t]{c}topos\\\mbox{ }\mbox{ }\end{tabular}}}
\put(0,0){\begin{picture}(200,0)(0,0)
\put(-50,-30){\makebox(50,20)[r]{\scriptsize$0\,$}}
\put(52,-30){\makebox(50,20)[r]{\scriptsize$1\,$}}
\put(150,-30){\makebox(50,20)[r]{\scriptsize$2\,$}}
\multiput(0,0)(100,0){2}{\begin{picture}(100,0)(0,0)
\put(0,0){\line(1,0){100}}
\put(0,0){\line(0,-1){30}}
\put(25,0){\line(0,-1){18}}
\put(50,0){\line(0,-1){24}}
\put(75,0){\line(0,-1){18}}
\put(100,0){\line(0,-1){30}}
\end{picture}}
\end{picture}}
\end{picture}
&
{\scriptsize \begin{tabular}{p{200pt}}
\\
\begin{itemize}
\item
Finite limits have been axiomatized and applied; 
exponents have been axiomatized and 
subobjects have been partialy axiomatized, 
but neither has yet been applied. 
\end{itemize} 
\end{tabular}}
\\
\setlength{\unitlength}{0.4pt}
\begin{picture}(200,0)(0,-20)
\put(25,10){\begin{picture}(0,25)(0,0)
\thicklines
\put(0,25){\circle*{5}}
\put(0,25){\vector(0,-1){25}}
\end{picture}}
\put(-120,-20){\makebox(50,0)[r]{\footnotesize\begin{tabular}[t]{c}{element}\end{tabular}}}
\put(-25,-60){\makebox(50,0){\tiny\begin{tabular}[t]{c}ordinary\end{tabular}}}
\put(75,-60){\makebox(50,0){\tiny\begin{tabular}[t]{c}generialized\end{tabular}}}
\put(0,0){\begin{picture}(200,0)(0,0)
\put(-50,-30){\makebox(50,20)[r]{\scriptsize$0\,$}}
\put(52,-30){\makebox(50,20)[r]{\scriptsize$1\,$}}
\multiput(0,0)(100,0){1}{\begin{picture}(100,0)(0,0)
\put(0,0){\line(1,0){100}}
\put(0,0){\line(0,-1){30}}
\put(25,0){\line(0,-1){18}}
\put(50,0){\line(0,-1){24}}
\put(75,0){\line(0,-1){18}}
\put(100,0){\line(0,-1){30}}
\end{picture}}
\end{picture}}
\end{picture}
&
{\scriptsize \begin{tabular}{p{200pt}}
\\
\begin{itemize}
\item
The belonging, inclusion and membership relations for generalized elements (morphisms) have been defined. But only some generalized elements have been explicitly used in place of ordinary (global) elements;
some generalized elements show up as parameters.
\end{itemize} 
\end{tabular}}
\end{tabular}
\end{center}
\caption{Current Development State}
\label{current:development:state}
\end{figure}
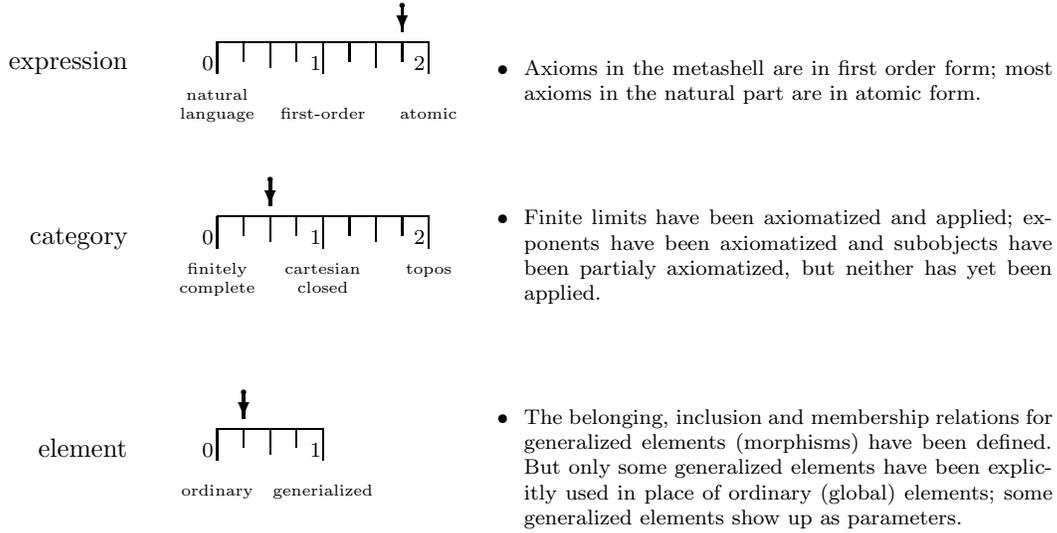

The IFF development is constantly under revision.
Attention and activity has moved 
from applications of institution theory to a category theory standard.
Several concepts about development have emerged:
the two design principles;
an architectural framework; and
three concurrent development processes
(Figure~\ref{current:development:state} indicates degree of completion):
axiomatic expression (natural language $\Rightarrow$ first order $\Rightarrow$ atomic);
category (finitely complete $\Rightarrow$ cartesian-closed $\Rightarrow$ topos); and
element (ordinary $\Rightarrow$ generalized).

\section{Architecture}\label{sec-architecture}

\begin{figure}
\begin{center}
\begin{tabular}{c}
\setlength{\unitlength}{0.55pt}
\begin{picture}(300,225)(-30,0)
\put(0,190){\begin{picture}(50,50)(0,0)
\qbezier[10](0,0)(25,0)(50,0)
\qbezier[10](0,0)(0,25)(0,50)
\qbezier[10](50,0)(25,25)(0,50)
\put(-48,22){\makebox(25,15)[r]{{\scriptsize\sffamily supra-natural}}}
\put(-48,8){\makebox(25,15)[r]{{\scriptsize\sffamily (metashell)}}}
\qbezier[10](-12.5,0)(-12.5,25)(-12.5,50)
\qbezier[1](-12.5,50)(-10,50)(-7.5,50)
\qbezier[1](-12.5,0)(-10,0)(-7.5,0)
\end{picture}}
\put(0,50){\begin{picture}(50,100)(0,0)
\put(0,0){\framebox(50,100){}}
\put(-48,52){\makebox(25,15)[r]{{\scriptsize\sffamily natural}}}
\put(-48,38){\makebox(25,15)[r]{{\scriptsize\sffamily (metalevel)}}}
\put(-12.5,0){\line(0,1){100}}
\put(-12.5,100){\line(1,0){5}}
\put(-12.5,0){\line(1,0){5}}
\end{picture}}
\put(54,50){\begin{picture}(50,100)(0,0)
\put(0,0){\framebox(50,100){}}
\end{picture}}
\put(125,50){\begin{picture}(50,100)(0,0)
\put(0,0){\line(0,1){100}}
\put(0,0){\line(1,0){100}}
\put(0,100){\line(1,-1){100}}
\end{picture}}
\put(0,0){\begin{picture}(250,20)(0,0)
\put(0,20){\line(1,0){225}}
\put(0,0){\line(1,0){245}}
\put(245,0){\line(-1,1){20}}
\put(0,0){\line(0,1){20}}
\put(-48,9.5){\makebox(25,15)[r]{{\scriptsize\sffamily objective}}}
\put(-48,-4.5){\makebox(25,15)[r]{{\scriptsize\sffamily (object level)}}}
\put(-12.5,0){\line(0,1){20}}
\put(-12.5,20){\line(1,0){5}}
\put(-12.5,0){\line(1,0){5}}
\end{picture}}
\put(12.5,160){\makebox(25,15){{\scriptsize\sffamily core}}}
\put(0,160){\line(1,0){50}}
\put(0,160){\line(0,-1){5}}
\put(50,160){\line(0,-1){5}}
\put(66.5,160){\makebox(25,15){{\scriptsize\sffamily structure}}}
\put(54,160){\line(1,0){50}}
\put(54,160){\line(0,-1){5}}
\put(104,160){\line(0,-1){5}}
\put(37.5,24){\makebox(25,15){{\scriptsize\sffamily pure}}}
\put(0,40){\line(1,0){104}}
\put(0,40){\line(0,1){5}}
\put(104,40){\line(0,1){5}}
\put(162.5,24){\makebox(25,15){{\scriptsize\sffamily applied}}}
\put(125,40){\line(1,0){100}}
\put(125,40){\line(0,1){5}}
\put(225,40){\line(0,1){5}}
\end{picture}
\\ \\
(exploded iconic version)
\\ \\
\setlength{\unitlength}{0.85pt}
\begin{picture}(357,225)(0,0)
\put(-40,165){\begin{picture}(10,45)(0,0)
\put(-55,15){\makebox(25,15){{\footnotesize\sffamily metashell}}}
\put(-55,3){\makebox(25,15){{\scriptsize (first order)}}}
\qbezier[1](-12.5,45)(-10,45)(-7.5,45)
\qbezier[9](-12.5,0)(-12.5,22.5)(-12.5,45)
\qbezier[1](-12.5,0)(-10,0)(-7.5,0)
\end{picture}}
\put(-40,30){\begin{picture}(10,105)(0,0)
\put(-55,43){\makebox(25,15){{\footnotesize\sffamily natural}}}
\put(-55,31){\makebox(25,15){{\footnotesize\sffamily part}}}
\put(-55,19){\makebox(25,15){{\scriptsize (atomic)}}}
\put(-12.5,90){\line(1,0){5}}
\put(-12.5,0){\line(0,1){90}}
\put(-12.5,0){\line(1,0){5}}
\end{picture}}
\put(0,165){\begin{picture}(60,45)(0,0)
 \qbezier[3](0,45)(7.5,45)(15,45)
 \qbezier[6](0,30)(15,30)(30,30)
\qbezier[12](0,15)(30,15)(60,15)
\qbezier[12](0,0)(30,0)(60,0)
\qbezier[3](15,30)(15,37.5)(15,45)
\qbezier[3](30,15)(30,22.5)(30,30)
\qbezier[3](60,0)(60,7.5)(60,15)
\qbezier[9](0,0)(0,22.5)(0,45)
\put(-32,30){\makebox(25,15)[r]{
{\tiny\sffamily iff}}
\makebox[0pt][l]{\scriptsize \begin{tabular}{@{\hspace{80pt}}p{139pt}}
\\
metashell: 
temporary scaffolding used 
to construct the metastack in particular and the natural part in general
\end{tabular}}}
\put(-32,15){\makebox(25,15)[r]{{\tiny\sffamily type}}}
\put(-32,0){\makebox(25,15)[r]{{\tiny\sffamily meta}}}
\end{picture}}
\put(0,30){\begin{picture}(357,135)(0,0)
\put(-30,0){\begin{picture}(30,135)(0,0)
\put(-2,75){\makebox(25,15)[r]{{\footnotesize $\mathbf{\cdots}$}}}
\put(-2,60){\makebox(25,15)[r]{{\tiny\sffamily n}}}
\put(-2,45){\makebox(25,15)[r]{{\footnotesize $\mathbf{\cdots}$}}}
\put(-2,30){\makebox(25,15)[r]{{\tiny\sffamily vlrg = 3}}}
\put(-2,15){\makebox(25,15)[r]{{\tiny\sffamily lrg = 2}}}
\put(-2,0){\makebox(25,15)[r]{{\tiny\sffamily sml = 1}}}
\put(-2,-25){\makebox(25,15)[r]{{\tiny\sffamily obj = 0}}}
\end{picture}}
\put(0,0){\begin{picture}(60,135)(0,0)
\put(0,0){\line(0,1){105}}
\put(60,0){\line(0,1){105}}
\put(0,105){\makebox(60,15){{\scriptsize\sffamily IFF-CORE}}}
\put(0,0){\begin{picture}(15,105)(0,0)
\put(0,90){\makebox(15,15){{\tiny\sffamily krnl}}}
\put(0,60){\framebox(15,15){\framebox(5,5){}}}
\put(0,30){\framebox(15,15){\framebox(5,5){}}}
\put(0,15){\framebox(15,15){\framebox(5,5){}}}
\put(0,0){\framebox(15,15){\rule{5pt}{5pt}}}
\put(7.5,50){\oval(20,112)}
\put(-125,107){\makebox(50,15){\scriptsize\sffamily{metastack}}}
\put(-100,114){\oval(50,18)}
\qbezier(-74.5,114)(3,114)(4.5,105.5)
\end{picture}}
\put(15,0){\begin{picture}(15,105)(0,0)
\put(0,90){\makebox(15,15){{\tiny\sffamily dgm}}}
\put(0,60){\framebox(15,15){\framebox(5,5){}}}
\put(0,30){\framebox(15,15){\framebox(5,5){}}}
\put(0,15){\framebox(15,15){\framebox(5,5){}}}
\put(0,0){\framebox(15,15){\rule{5pt}{5pt}}}
\end{picture}}
\put(30,0){\begin{picture}(15,105)(0,0)
\put(0,90){\makebox(15,15){{\tiny\sffamily lim}}}
\put(0,60){\framebox(15,15){\framebox(5,5){}}}
\put(0,30){\framebox(15,15){\framebox(5,5){}}}
\put(0,15){\framebox(15,15){\framebox(5,5){}}}
\put(0,0){\framebox(15,15){\rule{5pt}{5pt}}}
\end{picture}}
\put(45,0){\begin{picture}(15,105)(0,0)
\put(0,90){\makebox(15,15){{\tiny\sffamily expn}}}
\put(0,60){\framebox(15,15){\framebox(5,5){}}}
\put(0,30){\framebox(15,15){\framebox(5,5){}}}
\put(0,15){\framebox(15,15){\framebox(5,5){}}}
\put(0,0){\framebox(15,15){\rule{5pt}{5pt}}}
\end{picture}}
\end{picture}}
\put(62,0){\begin{picture}(105,135)(0,0)
\put(0,0){\line(0,1){105}}
\put(105,0){\line(0,1){105}}
\put(0,105){\makebox(105,15){{\scriptsize\sffamily IFF-CAT}}}
\put(0,0){\begin{picture}(15,105)(0,0)
\put(0,90){\makebox(15,15){{\tiny\sffamily gph}}}
\put(0,60){\framebox(15,15){\framebox(5,5){}}}
\put(0,30){\framebox(15,15){\framebox(5,5){}}}
\put(0,15){\framebox(15,15){\rule{5pt}{5pt}}}
\put(0,0){\framebox(15,15){\framebox(5,5){}}}
\end{picture}}
\put(15,0){\begin{picture}(15,105)(0,0)
\put(0,90){\makebox(15,15){{\tiny\sffamily cat}}}
\put(0,60){\framebox(15,15){\framebox(5,5){}}}
\put(0,30){\framebox(15,15){\framebox(5,5){}}}
\put(0,15){\framebox(15,15){\rule{5pt}{5pt}}}
\put(0,0){\framebox(15,15){\framebox(5,5){}}}
\end{picture}}
\put(30,0){\begin{picture}(15,105)(0,0)
\put(0,90){\makebox(15,15){{\tiny\sffamily func}}}
\put(0,60){\framebox(15,15){\framebox(5,5){}}}
\put(0,30){\framebox(15,15){\framebox(5,5){}}}
\put(0,15){\framebox(15,15){\rule{5pt}{5pt}}}
\put(0,0){\framebox(15,15){\framebox(5,5){}}}
\end{picture}}
\put(45,0){\begin{picture}(15,105)(0,0)
\put(0,90){\makebox(15,15){{\tiny\sffamily nat}}}
\put(0,60){\framebox(15,15){\framebox(5,5){}}}
\put(0,30){\framebox(15,15){\framebox(5,5){}}}
\put(0,15){\framebox(15,15){\rule{5pt}{5pt}}}
\put(0,0){\framebox(15,15){\framebox(5,5){}}}
\end{picture}}
\put(60,0){\begin{picture}(15,105)(0,0)
\put(0,90){\makebox(15,15){{\tiny\sffamily adj}}}
\put(0,60){\framebox(15,15){\framebox(5,5){}}}
\put(0,30){\framebox(15,15){\framebox(5,5){}}}
\put(0,15){\framebox(15,15){\rule{5pt}{5pt}}}
\put(0,0){\framebox(15,15){\framebox(5,5){}}}
\end{picture}}
\put(75,0){\begin{picture}(15,105)(0,0)
\put(0,90){\makebox(15,15){{\tiny\sffamily mnd}}}
\put(0,60){\framebox(15,15){\framebox(5,5){}}}
\put(0,30){\framebox(15,15){\framebox(5,5){}}}
\put(0,15){\framebox(15,15){\rule{5pt}{5pt}}}
\put(0,0){\framebox(15,15){\framebox(5,5){}}}
\end{picture}}
\put(90,0){\begin{picture}(15,105)(0,0)
\put(0,90){\makebox(15,15){{\tiny\sffamily kan}}}
\put(0,60){\framebox(15,15){\framebox(5,5){}}}
\put(0,30){\framebox(15,15){\framebox(5,5){}}}
\put(0,15){\framebox(15,15){\rule{5pt}{5pt}}}
\put(0,0){\framebox(15,15){\framebox(5,5){}}}
\end{picture}}
\end{picture}}
\put(169,0){\begin{picture}(15,135)(0,0)
\put(0,0){\line(0,1){105}}
\put(15,0){\line(0,1){105}}
\put(0,0){\begin{picture}(15,105)(0,0)
\put(0,90){\makebox(15,15){{\tiny\sffamily top}}}
\put(0,60){\framebox(15,15){\framebox(5,5){}}}
\put(0,30){\framebox(15,15){\framebox(5,5){}}}
\put(0,15){\framebox(15,15){\rule{5pt}{5pt}}}
\put(0,0){\framebox(15,15){\framebox(5,5){}}}
\end{picture}}
\end{picture}}
\put(186,0){\begin{picture}(15,135)(0,0)
\put(0,0){\line(0,1){105}}
\put(15,0){\line(0,1){105}}
\put(0,0){\begin{picture}(15,105)(0,0)
\put(0,90){\makebox(15,15){{\tiny\sffamily fbr}}}
\put(0,60){\framebox(15,15){\framebox(5,5){}}}
\put(0,30){\framebox(15,15){\framebox(5,5){}}}
\put(0,15){\framebox(15,15){\rule{5pt}{5pt}}}
\put(0,0){\framebox(15,15){\framebox(5,5){}}}
\end{picture}}
\end{picture}}
\put(203,0){\begin{picture}(60,135)(0,0)
\put(0,0){\line(0,1){105}}
\put(60,0){\line(0,1){105}}
\put(0,105){\makebox(60,15){{\scriptsize\sffamily IFF-INS}}}
\put(0,0){\begin{picture}(15,105)(0,0)
\put(0,90){\makebox(15,15){{\tiny\sffamily clsn}}}
\put(0,15){\framebox(15,15){\rule{5pt}{5pt}}}
\put(0,0){\framebox(15,15){\framebox(5,5){}}}
\end{picture}}
\put(15,0){\begin{picture}(15,105)(0,0)
\put(0,90){\makebox(15,15){{\tiny\sffamily info}}}
\put(0,15){\framebox(15,15){\rule{5pt}{5pt}}}
\put(0,0){\framebox(15,15){\framebox(5,5){}}}
\end{picture}}
\put(30,0){\begin{picture}(15,105)(0,0)
\put(0,90){\makebox(15,15){{\tiny\sffamily clg}}}
\put(0,15){\framebox(15,15){\rule{5pt}{5pt}}}
\put(0,0){\framebox(15,15){\framebox(5,5){}}}
\end{picture}}
\put(45,0){\begin{picture}(15,105)(0,0)
\put(0,90){\makebox(15,15){{\tiny\sffamily inst}}}
\put(0,15){\framebox(15,15){\rule{5pt}{5pt}}}
\put(0,0){\framebox(15,15){\framebox(5,5){}}}
\end{picture}}
\end{picture}}
\put(265,0){\begin{picture}(45,135)(0,0)
\put(0,0){\line(0,1){105}}
\put(45,0){\line(0,1){105}}
\put(0,105){\makebox(45,15){{\scriptsize\sffamily IFF-FOL}}}
\put(0,0){\begin{picture}(15,105)(0,0)
\put(0,90){\makebox(15,15){{\tiny\sffamily trm}}}
\put(0,0){\framebox(15,15){\rule{5pt}{5pt}}}
\end{picture}}
\put(15,0){\begin{picture}(15,105)(0,0)
\put(0,90){\makebox(15,15){{\tiny\sffamily expr}}}
\put(0,0){\framebox(15,15){\rule{5pt}{5pt}}}
\end{picture}}
\put(30,0){\begin{picture}(15,105)(0,0)
\put(0,90){\makebox(15,15){{\tiny\sffamily fol}}}
\put(0,0){\framebox(15,15){\rule{5pt}{5pt}}}
\end{picture}}
\end{picture}}
\put(312,0){\begin{picture}(45,135)(0,0)
\put(0,0){\line(0,1){105}}
\put(45,0){\line(0,1){105}}
\put(0,105){\makebox(45,15){{\scriptsize\sffamily IFF-OBJ}}}
\put(0,0){\begin{picture}(15,105)(0,0)
\put(0,90){\makebox(15,15){{\tiny\sffamily lang}}}
\put(0,0){\framebox(15,15){\rule{5pt}{5pt}}}
\end{picture}}
\put(15,0){\begin{picture}(15,105)(0,0)
\put(0,90){\makebox(15,15){{\tiny\sffamily mod}}}
\put(0,0){\framebox(15,15){\rule{5pt}{5pt}}}
\end{picture}}
\put(30,0){\begin{picture}(15,105)(0,0)
\put(0,90){\makebox(15,15){{\tiny\sffamily th}}}
\put(0,0){\framebox(15,15){\rule{5pt}{5pt}}}
\end{picture}}
\end{picture}}
\end{picture}}
\put(0,-3){\begin{picture}(357,30)(0,0)
\put(0,15){\begin{picture}(357,15)(0,0)
\put(0,0){\framebox(357,15){}}
\put(0,0){\begin{picture}(30,15)(0,0)
\put(0,0){\framebox(30,15){{\scriptsize\sffamily SUO}}}
\end{picture}}
\put(30,0){\begin{picture}(40,15)(0,0)
\put(0,0){\framebox(40,15){{\scriptsize\sffamily Botany}}}
\end{picture}}
\put(70,0){\begin{picture}(60,15)(0,0)
\put(0,0){\framebox(60,15){{\scriptsize\sffamily Ontolingua}}}
\end{picture}}
\put(130,0){\begin{picture}(52,15)(0,0)
\put(0,0){\framebox(52,15){{\scriptsize\sffamily WordNet}}}
\end{picture}}
\put(182,0){\begin{picture}(50,15)(0,0)
\put(0,0){\framebox(50,15){{\scriptsize\sffamily SENSUS}}}
\end{picture}}
\put(232,0){\begin{picture}(40,15)(0,0)
\put(0,0){\framebox(40,15){{\scriptsize\sffamily Holes}}}
\end{picture}}
\put(272,0){\begin{picture}(40,15)(0,0)
\put(0,0){\framebox(40,15){{\scriptsize\sffamily Gene}}}
\end{picture}}
\put(312,0){\begin{picture}(45,15)(0,0)
\put(0,0){\framebox(45,15){{\scriptsize\sffamily SemWeb}}}
\end{picture}}
\end{picture}}
\put(0,0){\begin{picture}(357,15)(0,0)
\put(0,0){\framebox(357,15){}}
\put(0,0){\begin{picture}(23,15)(0,0)
\put(0,0){\framebox(23,15){{\scriptsize\sffamily Cyc}}}
\end{picture}}
\put(23,0){\begin{picture}(47,15)(0,0)
\put(0,0){\framebox(47,15){{\scriptsize\sffamily Enterprise}}}
\end{picture}}
\put(70,0){\begin{picture}(60,15)(0,0)
\put(0,0){\framebox(60,15){{\scriptsize\sffamily e-commerce}}}
\end{picture}}
\put(130,0){\begin{picture}(52,15)(0,0)
\put(0,0){\framebox(52,15){{\scriptsize\sffamily Government}}}
\end{picture}}
\put(182,0){\begin{picture}(50,15)(0,0)
\put(0,0){\framebox(50,15){{\scriptsize\sffamily Education}}}
\end{picture}}
\put(232,0){\begin{picture}(40,15)(0,0)
\put(0,0){\framebox(40,15){{\scriptsize\sffamily HPKB}}}
\end{picture}}
\put(272,0){\begin{picture}(40,15)(0,0)
\put(0,0){\framebox(40,15){{\scriptsize\sffamily SUMO}}}
\end{picture}}
\put(312,0){\begin{picture}(45,15)(0,0)
\put(0,0){\framebox(45,15){{\footnotesize $\mathbf{\cdots}$}}}
\end{picture}}
\end{picture}}
\end{picture}}
\end{picture}
\\ \\
(detailed version)
\\
\\
\end{tabular}
\end{center}
\caption{The IFF Architecture}
\label{architecture}
\end{figure}

\begin{flushleft}
{\small \begin{tabular}{p{4in}}
\emph{``Such is `set theory' in the practice of mathematics;}       \\
\emph{it is part of the essence from which organization emerges.''} \\
$\sim$ F.W.~Lawvere \rule{0pt}{10pt}
\end{tabular}}
\end{flushleft}
The new IFF architecture (Figure~\ref{architecture})
--- a two dimensional structure consisting of 
levels (the vertical dimension), 
namespaces (the horizontal dimension) and 
meta-ontologies (coherent composites of namespaces)
--- is described in terms of parts, aspects and components. 
The vertical dimension of the IFF Architecture consists of three parts:
the objective part ($n = 0$),
the natural part ($1 \leq n < \infty$) and
the supranatural part 
({\footnotesize $n \in \{ \mathsf{meta}, \mathsf{type}, \mathsf{iff} \}$}).
Along the horizontal dimension, 
the natural part is partitioned into pure and applied aspects, and 
the pure aspect is partitioned into core and structural components. 
The pure aspect,
which contains meta-ontologies axiomatizing in adjunctive form
the set-theoretic and category-theoretic foundations needed elsewhere in the IFF, 
is the first step towards a standard ontology for category theory.
The kernel of the core component, called the IFF metastack, 
represents a chain of toposes\footnote{A category $\mathcal{E}$ is a topos when
it has finite limits,
it is cartesian closed, and
it has a subobject classifier;
equivalently, when
it has finite limits and comes equipped with
an object of truth values $\Omega_{\mathcal{E}} \in \mathsf{obj}(\mathcal{E})$,
a power function 
$\mbox{{\large $\wp$}}_{\mathcal{E}} 
:  \mathsf{obj}(\mathcal{E}) \rightarrow \mathsf{obj}(\mathcal{E})$
where $\mathsf{Sub}_{\mathcal{E}}A$ is the set of subobjects of $A$,
and two natural isomorphisms
$\mathsf{Sub}_{\mathcal{E}}(A) \cong \mathsf{Hom}_{\mathcal{E}}[A,\Omega_{\mathcal{E}}]$
and
$\mathsf{Hom}_{\mathcal{E}}[A{\times}B,\Omega_{\mathcal{E}}]
\cong \mathsf{Hom}_{\mathcal{E}}[A,\mbox{{\large $\wp$}}_{\mathcal{E}}B]$.}
$\mathsf{Set} 
= \langle \mathsf{Set}_1 \subset \mathsf{Set}_2 \subset \cdots \subset \mathsf{Set}_n \subset \cdots \rangle$
anchoring the entire IFF architecture. 
The metastack structure is based upon Cantor's diagonal argument. 
The applied aspect contains meta-ontologies providing 
the terminology and axiomatization needed for 
the logical and semiotic functionality in applications. 

\subsection{Modular Structure}\label{subsec-modular:structure}

The modular structure of the IFF architecture 
consists of parts, aspects and components.

\subsubsection{Metashell}\label{subsubsec-metashell}

The supranatural part is also called the IFF metashell.
The form of the metashell axiomatization 
has a first order expression with explicit logical structure.
The content of the metashell axiomatization 
consists of (only) one namespace at each of the three metalevels
{\footnotesize $\{ \mathsf{meta}, \mathsf{type}, \mathsf{iff} \}$}:
the $\mathsf{iff}$  level contains the {\small\ttfamily IFF-IFF} namespace,
the $\mathsf{type}$ level contains the {\small\ttfamily IFF-TYPE} namespace, and
the $\mathsf{meta}$ level contains the {\small\ttfamily IFF-META} namespace.
The {\small\ttfamily IFF-IFF} namespace
axiomatizes a directed graph of abstract sets and functions\footnote{It 
contains only the five terms
$\{\mbox{`{\scriptsize\ttfamily thing}'},
\mbox{`{\scriptsize\ttfamily set}'},
\mbox{`{\scriptsize\ttfamily function}'},
\mbox{`{\scriptsize\ttfamily source}'},
\mbox{`{\scriptsize\ttfamily target}'}\}$.},
the {\small\ttfamily IFF-TYPE} namespace
axiomatizes a finitely-complete category of abstract sets and functions\footnote{It
contains approximately 450 terms,
with 250 terms in the kernel subnamespace
and 100 terms each in the diagam and limit subnamespaces.}, and
the {\small\ttfamily IFF-META} namespace 
axiomatizes a topos of Cantorian featureless abstract sets and functions.
From one point of view,
the metashell serves as a temporary scaffolding for construction of 
the entire IFF architecture in the natural part.

\subsubsection{Objective Part}\label{subsubsec-objective:part}

The form of the objective part axiomatization 
has a simple atomic expression with no logical structure.
The content of the objective part axiomatization 
consists of terminology for object-level ontologies.
The form of the natural part axiomatization 
also has a simple atomic expression with some implicit logical structure.
The content of the natural part axiomatization 
consists of namespaces representing many of the basic concepts of mathematics and logic.
By conforming to the principle of categorical design,
the logical expression in the natural part is atomic,
thus using commutative diagrams 
to express the intuitions and ideas of the working category theorist. 

\subsubsection{Pure Aspect}\label{subsubsec-pure:aspect}

\begin{table}
\begin{center}
{\scriptsize \begin{tabular}{r@{\hspace{5pt}$\in$\hspace{5pt}}l@{\hspace{15pt}}r@{\hspace{5pt}=\hspace{5pt}}l@{\hspace{15pt}}l}
\multicolumn{2}{c}{} & \multicolumn{2}{c}{$\cdots$} & \\
  $\mathsf{Set}_{n}$ & $\mathsf{obj}(\mathsf{Cat}_{n{+}1})$ 
& $\mathsf{Cat}_{n{+}1}$ & $\mathsf{cat}(\mathsf{Set}_{n{+}1})$ & level ${n{+}1}$ categories \\
  $\mathsf{Set}_{n-1}$ & $\mathsf{obj}(\mathsf{Cat}_n)$ 
& $\mathsf{Cat}_n$ & $\mathsf{cat}(\mathsf{Set}_n)$ & level $n$ categories \\
\multicolumn{2}{c}{} & \multicolumn{2}{c}{$\cdots$} & \\
  $\mathsf{Set}_1$ & $\mathsf{obj}(\mathsf{Cat}_2)$ 
& $\mathsf{Cat}_2$ & $\mathsf{cat}(\mathsf{Set}_2)$ & large categories     \\
\multicolumn{2}{c}{}
& $\mathsf{Cat}_1$ & $\mathsf{cat}(\mathsf{Set}_1)$ & small categories     \\
\end{tabular}}
\end{center}
\caption{Topos Chain}
\label{topos:chain}
\end{table}

The pure aspect of the IFF architecture 
is partitioned into a core component and a structural component.
The core component represents set theory.
The structural component represents category theory.
The set-theoretic and category-theoretic axiomatizations constrain each other:
any category is defined as a set of objects and a set of morphisms with connecting functions;
whereas the collections of sets and functions (at any metalevel) form a category.
The axiomatization for any concept in the pure aspect 
is given in one generic module (namespace) at level $n$, for $1 \leq n < \infty$.
The finite metalevels, $1 \leq n < \infty$, 
are populated by generic and parametric meta-ontologies.
Generic means that the terminology and axiomatization for any two metalevels is identical.
Parametric means that the metalevel index is a parameter.
Only one copy of a meta-ontology with a level parameter is needed for all finite levels.

\subsubsection{Core Component}\label{subsubsec-core:component}

The core component of the IFF architecture
contains a single generic level $n$ meta-ontology {\ttfamily\small IFF-SET} for set theory, 
which incorporates the specialization of 
either the level $n{+}1$ version of itself and/or
the {\small\ttfamily IFF-META} namespace from the metashell.
The {\small\ttfamily IFF-SET} meta-ontology specifies set theory as 
a chain of toposes (Table~\ref{topos:chain}) of Cantorian featureless abstract sets
\begin{center}
{\small $\begin{array}{l}
\mathsf{Set} = \left\langle
\mathsf{Set}_1 \subset \mathsf{Set}_2 \subset \cdots \subset \mathsf{Set}_n \subset \cdots
\right\rangle
\end{array}$}
\end{center}
where $\mathsf{Set}_1$ contains ``small'' sets and functions between ``small'' sets
and   $\mathsf{Set}_2$ contains ``large'' sets and functions between ``large'' sets.
This chain is motivated by and compatible with the Cantorian expansion of sets 
discussed in subsection~\ref{subsec-cantorian:expansion}.

\subsubsection{Structure Component}\label{subsubsec-structure:component}

The structure component of the IFF architecture
contains various generic meta-ontologies for category theory,
({\ttfamily\small IFF-CAT}, {\ttfamily\small IFF-2CAT}, {\ttfamily\small IFF-DCAT}, \ldots ).
By axiomatizing the basic concepts of category theory,
such as categories, functors, natural transformations, adjunctions and monads,
the {\small\ttfamily IFF-CAT} meta-ontology specifies category theory as a chain  (Table~\ref{topos:chain}) of internal categories
\begin{center}
{\small $\begin{array}{l}
\mathsf{Cat} = \left\langle
\mathsf{Cat}_1 \subset \mathsf{Cat}_2 \subset \cdots \subset \mathsf{Cat}_n \subset \cdots 
\right\rangle
\end{array}$}
\end{center}
in the toposes $\mathsf{Set}$.
The generic level axiomatizations 
for the selected topos-representing terms in Table~\ref{set:topos}
justify the assertion: 
for each $n$, 
the category $\mathsf{Set}_{n}$ of level $n$ sets and functions is a topos.

\begin{table}
\begin{center}
{\scriptsize \begin{tabular}{|ll|}
\multicolumn{1}{l}{\rule[-5pt]{0pt}{12pt}{\sffamily\normalsize IFF Term}} 
& \multicolumn{1}{l}{\sffamily\normalsize Concept}                                \\ \hline
\rule[-4pt]{0pt}{14pt}
{\scriptsize\ttfamily \#n:Set}                  & $\mathsf{Set}_n$                \\ \hline
\rule[-4pt]{0pt}{14pt}
{\scriptsize\ttfamily \#n.set:set}              & $\mathsf{obj}(\mathsf{Set}_n)$  \\
\rule[-4pt]{0pt}{10pt}
{\scriptsize\ttfamily \#n.set:power}            & $\mathsf{obj}(\mbox{{\normalsize $\wp$}}_{\mathsf{Set}_n})$ \\
\rule[-5pt]{0pt}{12pt}
{\scriptsize\ttfamily \#n.set:$\{$zero, one, two$\}$} 
& $0_{\mathsf{Set}_n}\!,
   1_{\mathsf{Set}_n}\!,
   2_{\mathsf{Set}_n} \!\!\!= \Omega_{\mathsf{Set}_n}$ \\ \hline
\rule[-4pt]{0pt}{10pt}
{\scriptsize\ttfamily \#n.ftn:function}         & $\mathsf{mor}(\mathsf{Set}_n)$  \\
\rule[-4pt]{0pt}{10pt}
{\scriptsize\ttfamily \#n.ftn:power}            & $\mathsf{mor}(\mbox{{\normalsize $\wp$}}_{\mathsf{Set}_n})$ \\
\rule[-4pt]{0pt}{10pt}
{\scriptsize\ttfamily \#n.ftn:composition}      & $\circ_{\mathsf{Set}_n}$        \\
\rule[-5pt]{0pt}{12pt}
{\scriptsize\ttfamily \#n.ftn:identity}         & $1_{\mathsf{Set}_n}$            \\ \hline
\rule[-4pt]{0pt}{14pt}
{\scriptsize\ttfamily \#n.pred:fiber}           & $\mathsf{Sub}_{\mathsf{Set}_n}$ \\
\rule[-5pt]{0pt}{12pt}
{\scriptsize\ttfamily \#n.pred:binary-meet}     & $\wedge_{\mathsf{Set}_n}$       \\ \hline
\rule[-5pt]{0pt}{15pt}
{\scriptsize\ttfamily \#n.rel:fiber01}          & $\varphi^{\mathsf{Set}_n}_{01}$ \\ \hline
\rule[-4pt]{0pt}{15pt}
{\scriptsize\ttfamily \#n.lim.prd2.obj:product} & $\!{\times}_{\mathsf{Set}_n}$   \\
\rule[-5pt]{0pt}{14pt}
{\scriptsize\ttfamily \#n.lim.prd2.obj:projection$\{$0,1$\}$} 
& ${\pi}_i^{\mathsf{Set}_n}\!\!\!\!\!\!\!,\;\;\;\;\;\; i = 0,1$               \\ \hline
\rule[-4pt]{0pt}{14pt}
{\scriptsize\ttfamily \#n.exp.obj:exponent}     & $B^A$                          \\
\rule[-4pt]{0pt}{10pt}
{\scriptsize\ttfamily \#n.exp.obj:evaluation}   & $B^A {\times} A \rightarrow B$ \\
\rule[-4pt]{0pt}{10pt}
{\scriptsize\ttfamily \#n.exp.obj:hom}          & ${\mathsf{Set}_n}[\mbox{-},\mbox{-}]$ \\ 
\rule[-5pt]{0pt}{12pt}
{\scriptsize\ttfamily \#n.exp.obj:curry}      
& ${\mathsf{Set}_n}[C{\times}A,B] \rightarrow {\mathsf{Set}_n}[C,B^A]$            \\ \hline
\end{tabular}}
\end{center}
\caption{$\mathsf{Set}_n$ as a Topos}
\label{set:topos}
\end{table}

\subsubsection{Atomic Expression}\label{subsubsec-atomic:expression}

To reiterate,
the natural part of the IFF is where most of
the concepts and constraints of mathematics and logic are represented.
This representation is atomic,
consisting of either declarations or equations.
The declarations state that 
an element is in a set or 
a pair of elements satisfies a relation.
The equations, 
which have only unary functions in their terms, 
are equivalent to the commutative diagrams of category theory.
Hence,
in the IFF all of the axiomatizations of category theory, set theory, logic, institution theory, science, et cetera, are in the form of declarations or commutative diagrams.
This is in contrast to the usual notion 
that category-theoretic representations are first order.

\subsection{Other Features}\label{subsec-other:features}

Other features of the IFF architecture include:
the meta\-stack, 
a simplification that erases links to the metashell,
the transition from a previous architecture \cite{kent:04},
the IFF grammar, and
the IFF as hierarchical metalanguage. 

\subsubsection{Metastack}\label{subsubsec-metastack}

The IFF metastack,
which is the kernel of the core component,
represents the Cantorian expansion.
It binds and anchors the natural part of the IFF, 
and connects the natural part to the metashell.
The metastack contains a lattice-like structure 
for the fundamental subset ($\subseteq$), restriction ($\sqsubseteq$) and optimal-restriction ($\dot{\sqsubseteq}$), delimitation ($\leq$) and abridgment ($\preceq$) orders
on sets, functions, predicates and relations,
respectively.
This is illustrated in Figure~\ref{kernel:chains},
where
$\partial_i$ are the source and target for functions with pairing $(\partial_i)$,
$\gamma$ and $\delta$ are the genus and differentia for predicates,
$\sigma_i$ are the component sets for relations with pairing $(\sigma_i)$,
and
$\varepsilon$ is the extent function for relations.
Just as 
(binary) relations are predicates (unary relations or parts) on a binary product and predicates are special functions (injections),
so also abridgment is a special case of delimitation and delimitaton is a special case of optimal-restriction.

\begin{table}
\begin{center}
\begin{tabular}{c}
\begin{tabular}{l}
{\scriptsize $\begin{array}{c@{\hspace{8pt}\subseteq\hspace{8pt}}c@{\hspace{8pt}\subseteq\hspace{8pt}}c@{\hspace{8pt}\subseteq\hspace{8pt}}c@{\hspace{8pt}\subseteq\hspace{8pt}}c@{\hspace{8pt}\subseteq\hspace{8pt}}c@{\hspace{8pt}\subseteq\hspace{8pt}}c}
\cdots & \mathsf{set}_{n} & \mathsf{set}_{n{+}1} & 
\cdots & \mathsf{set}_{\scriptscriptstyle \mathsf{meta}} & \mathsf{set}_{\scriptscriptstyle \mathsf{type}} & \mathsf{set}_{\scriptscriptstyle \mathsf{iff}}
\\
\cdots & \mathsf{ftn}_{n} & \mathsf{ftn}_{n{+}1} & 
\cdots & \mathsf{ftn}_{\scriptscriptstyle \mathsf{meta}} & \mathsf{ftn}_{\scriptscriptstyle \mathsf{type}} & \mathsf{ftn}_{\scriptscriptstyle \mathsf{iff}}
\\
\cdots & \mathsf{pred}_{n} & \mathsf{pred}_{n{+}1} & 
\cdots & \mathsf{pred}_{\scriptscriptstyle \mathsf{meta}} & \mathsf{pred}_{\scriptscriptstyle \mathsf{type}} & \mathsf{pred}_{\scriptscriptstyle \mathsf{iff}}
\\
\cdots & \mathsf{rel}_{n} & \mathsf{rel}_{n{+}1} & 
\cdots & \mathsf{rel}_{\scriptscriptstyle \mathsf{meta}} & \mathsf{rel}_{\scriptscriptstyle \mathsf{type}} & \mathsf{rel}_{\scriptscriptstyle \mathsf{iff}}
\\
\cdots & \mathsf{endo}_{n} & \mathsf{endo}_{n{+}1} & 
\cdots & \mathsf{endo}_{\scriptscriptstyle \mathsf{meta}} & \mathsf{endo}_{\scriptscriptstyle \mathsf{type}} & \mathsf{endo}_{\scriptscriptstyle \mathsf{iff}}
\end{array}$}
\end{tabular}
\\ 
{\footnotesize Subset} \rule{0pt}{15pt}
\\ \\
\begin{tabular}{l}
{\scriptsize $\begin{array}{c@{\hspace{8pt}\sqsubseteq\hspace{8pt}}c@{\hspace{8pt}\sqsubseteq\hspace{8pt}}c@{\hspace{8pt}\sqsubseteq\hspace{8pt}}c@{\hspace{8pt}\sqsubseteq\hspace{8pt}}c@{\hspace{8pt}\sqsubseteq\hspace{8pt}}c@{\hspace{8pt}\sqsubseteq\hspace{8pt}}c}
\cdots & \partial_{\scriptscriptstyle 0}^{n} & \partial_{\scriptscriptstyle 0}^{n{+}1} & 
\cdots & \partial_{\scriptscriptstyle 0}^{\scriptscriptstyle \mathsf{meta}} & \partial_{\scriptscriptstyle 0}^{\scriptscriptstyle \mathsf{type}} & \partial_{\scriptscriptstyle 0}^{\scriptscriptstyle \mathsf{iff}}
\\
\cdots & \partial_{\scriptscriptstyle 1}^{n} & \partial_{\scriptscriptstyle 1}^{n{+}1} & 
\cdots & \partial_{\scriptscriptstyle 1}^{\scriptscriptstyle \mathsf{meta}} & \partial_{\scriptscriptstyle 1}^{\scriptscriptstyle \mathsf{type}} & \partial_{\scriptscriptstyle 1}^{\scriptscriptstyle \mathsf{iff}}
\\
\cdots & {\delta}_{n} & {\delta}_{n{+}1} & 
\cdots & {\delta}_{\scriptscriptstyle \mathsf{meta}} 
& {\delta}_{\scriptscriptstyle \mathsf{type}} & {\delta}_{\scriptscriptstyle \mathsf{iff}}
\\
\cdots & {\varepsilon}_{n} & {\varepsilon}_{n{+}1} & 
\cdots & {\varepsilon}_{\scriptscriptstyle \mathsf{meta}} 
& {\varepsilon}_{\scriptscriptstyle \mathsf{type}} & {\varepsilon}_{\scriptscriptstyle \mathsf{iff}}
\\
\cdots & \sigma_{\scriptscriptstyle 0}^{n} & \sigma_{\scriptscriptstyle 0}^{n{+}1} & 
\cdots & \sigma_{\scriptscriptstyle 0}^{\scriptscriptstyle \mathsf{meta}} & \sigma_{\scriptscriptstyle 0}^{\scriptscriptstyle \mathsf{type}} & \sigma_{\scriptscriptstyle 0}^{\scriptscriptstyle \mathsf{iff}}
\\
\cdots & \sigma_{\scriptscriptstyle 1}^{n} & \sigma_{\scriptscriptstyle 1}^{n{+}1} & 
\cdots & \sigma_{\scriptscriptstyle 1}^{\scriptscriptstyle \mathsf{meta}} & \sigma_{\scriptscriptstyle 1}^{\scriptscriptstyle \mathsf{type}} & \sigma_{\scriptscriptstyle 1}^{\scriptscriptstyle \mathsf{iff}}
\end{array}$}
\\
{\scriptsize $\begin{array}{c@{\hspace{7.5pt}\dot{\sqsubseteq}\hspace{3.5pt}}c@{\hspace{3.5pt}\dot{\sqsubseteq}\hspace{3.5pt}}c@{\hspace{3.5pt}\dot{\sqsubseteq}\hspace{3.5pt}}c@{\hspace{3.5pt}\dot{\sqsubseteq}\hspace{3.5pt}}c@{\hspace{3.5pt}\dot{\sqsubseteq}\hspace{3.5pt}}c@{\hspace{3.5pt}\dot{\sqsubseteq}\hspace{3.5pt}}c}
\multicolumn{7}{l}{} \\
\cdots & (\partial_{\scriptscriptstyle i=0\!,\!1}^{n}) & (\partial_{\scriptscriptstyle i=0\!,\!1}^{n{+}1}) & 
\cdots & (\partial_{\scriptscriptstyle i=0\!,\!1}^{\scriptscriptstyle \mathsf{meta}}) & (\partial_{\scriptscriptstyle i=0\!,\!1}^{\scriptscriptstyle \mathsf{type}}) & (\partial_{\scriptscriptstyle i=0\!,\!1}^{\scriptscriptstyle \mathsf{iff}})
\\
\cdots & {\gamma}_{n} & {\gamma}_{n{+}1} & 
\cdots & {\gamma}_{\scriptscriptstyle \mathsf{meta}} 
& {\gamma}_{\scriptscriptstyle \mathsf{type}} & {\gamma}_{\scriptscriptstyle \mathsf{iff}}
\\
\cdots & (\sigma_{\scriptscriptstyle i=0\!,\!1}^{n}) & (\sigma_{\scriptscriptstyle i=0\!,\!1}^{n{+}1}) & 
\cdots & (\sigma_{\scriptscriptstyle i=0\!,\!1}^{\scriptscriptstyle \mathsf{meta}}) & (\sigma_{\scriptscriptstyle i=0\!,\!1}^{\scriptscriptstyle \mathsf{type}}) & (\sigma_{\scriptscriptstyle i=0\!,\!1}^{\scriptscriptstyle \mathsf{iff}})
\\
\end{array}$}
\end{tabular}
\\ {\footnotesize (Optimal-)Restriction} \rule{0pt}{15pt}
\\ \\
\begin{tabular}{l}
{\scriptsize $\begin{array}{c@{\hspace{8pt}\preceq\hspace{8pt}}c@{\hspace{8pt}\preceq\hspace{8pt}}c@{\hspace{8pt}\preceq\hspace{8pt}}c@{\hspace{8pt}\preceq\hspace{8pt}}c@{\hspace{8pt}\preceq\hspace{8pt}}c@{\hspace{8pt}\preceq\hspace{8pt}}c}
\cdots & {\subseteq}_{n} & {\subseteq}_{n{+}1} & 
\cdots & {\subseteq}_{\scriptscriptstyle \mathsf{meta}} & {\subseteq}_{\scriptscriptstyle \mathsf{type}} & {\subseteq}_{\scriptscriptstyle \mathsf{iff}}
\\
\cdots & {\leq}_{n} & {\leq}_{n{+}1} & 
\cdots & {\leq}_{\scriptscriptstyle \mathsf{meta}} & {\leq}_{\scriptscriptstyle \mathsf{type}} & {\leq}_{\scriptscriptstyle \mathsf{iff}}
\\
\cdots & {\sqsubseteq}_{n} & {\sqsubseteq}_{n{+}1} & 
\cdots & {\sqsubseteq}_{\scriptscriptstyle \mathsf{meta}} 
& {\sqsubseteq}_{\scriptscriptstyle \mathsf{type}} & {\sqsubseteq}_{\scriptscriptstyle \mathsf{iff}}
\\
\cdots & {\preceq}_{n} & {\preceq}_{n{+}1} & 
\cdots & {\preceq}_{\scriptscriptstyle \mathsf{meta}} & {\preceq}_{\scriptscriptstyle \mathsf{type}} & {\preceq}_{\scriptscriptstyle \mathsf{iff}}
\end{array}$}
\end{tabular}
\\ {\footnotesize Abridgment} \rule{0pt}{15pt}
\end{tabular}
\end{center}
\caption{Kernel Chains}
\label{kernel:chains}
\end{table}

\subsubsection{Simplification}\label{subsubsec-simplification}

Consider, amongst others, the level $n$ source function 
$\partial_0^n : \mathsf{ftn}_n \rightarrow \mathsf{set}_n$,
a level $n{+}1$ function.
We intend this to be a restriction of both 
the level $n{+}1$ source function 
$\partial_0^n \sqsubseteq \partial_0^{n{+}1}$
and the {\small\ttfamily meta} level source function
$\partial_0^n \sqsubseteq \partial_0^{\mathtt{meta}}$.
But since restriction defines the smaller function,
the latter link is clearly redundant.
The links, such as the latter, for the basic metastack orders
(subset, function (optimal-) restriction, predicate delimitation and relation abridgment)
between the natural part and the metashell are special.
These were used in the experimental phase of IFF development
to help develop the axiomatization for the natural part into a first order expression.
But as the principle of categorical design pushed 
towards an atomic expression for the natural part,
the links to the metashell were needed less and less.
With full atomic expression for the natural part,
it appears that we can make a choice whether to use these or not.
A minimal axiomatization would not need these.
It appears that with the full atomic expression 
they are only needed for added justification.
A set-theorist,
such as Feferman \cite{feferman:77},
who believes in the preeminence of 
the basic set-theoretic notions of ``collection'' and ``operation''
(represented in the IFF by the set and function collections in the metashell,
in particular those at the {\small\ttfamily iff} metalevel),
may want these links to remain in place.
But
the IFF is intended to have a category-theoretic foundation.
Thus,
we can choose whether or not to keep these links.
This is why we say 
the metashell is a temporary scaffolding used to construct the natural part.

\subsubsection{Transition}\label{subsubsec-transition}

\begin{figure}
\begin{center}
\begin{tabular}{c}
\\
\setlength{\unitlength}{0.65pt}
\begin{picture}(280,140)(0,-40)
\put(0,0){\begin{picture}(120,100)(0,0)
\put(60,85){\makebox(0,0){\scriptsize $\mathsf{ur}$}}
\put(60,70){\makebox(0,0){\scriptsize $\mathsf{top}$}}
\put(60,50){\makebox(0,0){\scriptsize $\mathsf{upper}$}}
\put(60,30){\makebox(0,0){\scriptsize $\mathsf{lower}$}}
\put(60,10){\makebox(0,0){\scriptsize $\mathsf{obj}$}}
\put(60,-20){\makebox(0,0){\footnotesize\sffamily old architecture}}
\put(0,0){\line(1,0){120}}
\put(0,0){\line(3,5){60}}
\put(120,0){\line(-3,5){60}}
\put(48,80){\line(1,0){24}}
\put(36,60){\line(1,0){48}}
\put(24,40){\line(1,0){72}}
\put(12,20){\line(1,0){96}}
\end{picture}}
\put(160,0){\begin{picture}(120,100)(0,0)
\put(60,85){\makebox(0,0){\scriptsize $\mathsf{iff}$}}
\put(60,70){\makebox(0,0){\scriptsize $\mathsf{type}$}}
\put(60,50){\makebox(0,0){\scriptsize $\mathsf{meta}$}}
\put(60,35){\makebox(0,0){\scriptsize $\mathsf{generic}$}}
\put(60,25){\makebox(0,0){\tiny (level $n$)}}
\put(60,10){\makebox(0,0){\scriptsize $\mathsf{obj}$}}
\put(60,-20){\makebox(0,0){\footnotesize\sffamily new architecture}}
\put(0,0){\line(1,0){120}}
\put(0,0){\line(3,5){60}}
\put(120,0){\line(-3,5){60}}
\put(48,80){\line(1,0){24}}
\put(36,60){\line(1,0){48}}
\put(24,40){\line(1,0){72}}
\put(12,20){\line(1,0){96}}
\end{picture}}
\put(140,90){\makebox(0,0){\small $\Rightarrow$}}
\put(140,70){\makebox(0,0){\small $\Rightarrow$}}
\put(140,50){\makebox(0,0){\small $\Rightarrow$}}
\put(140,30){\makebox(0,0){\small $\Rightarrow$}}
\multiput(72,80)(4,0){35}{\circle*{2}}
\multiput(84,60)(4,0){29}{\circle*{2}}
\multiput(96,40)(4,0){23}{\circle*{2}}
\multiput(108,20)(4,0){17}{\circle*{2}}
\put(315,71){\makebox(0,0){\scriptsize $\mathsf{metashell}$}}
\put(308,37){\makebox(0,0){\scriptsize $\mathsf{natural}$}}
\put(308,27){\makebox(0,0){\scriptsize $\mathsf{part}$}}
\put(276,96){\oval(8,8)[tr]}
\put(280,44){\line(0,1){22}}
\put(280,96){\line(0,-1){22}}
\put(276,44){\oval(8,8)[br]}
\put(284,74){\oval(8,8)[bl]}
\put(284,66){\oval(8,8)[tl]}
\put(276,36){\oval(8,8)[tr]}
\put(280,24){\line(0,1){2}}
\put(280,36){\line(0,-1){2}}
\put(276,24){\oval(8,8)[br]}
\put(284,34){\oval(8,8)[bl]}
\put(284,26){\oval(8,8)[tl]}
\end{picture}
\\ \\

{\scriptsize \begin{tabular}{rlp{80pt}@{\hspace{5pt}$\Leftarrow$\hspace{5pt}}l}
  {\sffamily part} 
& {\sffamily namespace} 
& \multicolumn{1}{l}{\sffamily contents} 
& \multicolumn{1}{@{\hspace{0pt}}l}{\sffamily contributors} \\
\multicolumn{3}{l}{} 
& \multicolumn{1}{@{\hspace{0pt}}l}{\ttfamily IFF-} \\
& iff 
& {\tiny basic sets, ftns}
& \hspace{15pt}{\ttfamily UR} \\
{\sffamily metashell}    
& type  
& $\mathsf{Set}_\mathsf{type},$ {\tiny category of sets}
& \hspace{15pt}{\ttfamily UR} \\
&   
& \tiny{\&} {\tiny spns, preds, rels}
& \hspace{15pt}{\ttfamily UR} \\
&   
& \tiny{\&} {\tiny finite limits} {\tiny (adj style)}
& \hspace{15pt}{\ttfamily TCO} \\
& meta
& $\mathsf{Set}_\mathsf{meta},$ {\tiny topos of sets}
& \hspace{15pt}{\ttfamily UCO} \\
&
& \tiny{\&} {\tiny exponents}, {\tiny (co)limits}
& \hspace{15pt}{\ttfamily UCO} \rule[-3pt]{0pt}{5pt} \\ \hline\hline
{\sffamily natural} 
& generic 
& $\mathsf{Set}_n,$ {\tiny topos of sets}
& \hspace{15pt}{\ttfamily LCO} 
\rule[3pt]{0pt}{5pt}
\\
&& \tiny{\&} {\tiny (everything else)}
& \hspace{15pt}{\ttfamily INS} 
\\
\multicolumn{3}{l}{} 
& \hspace{15.5pt}{\ttfamily FOL,ONT} 
\end{tabular}}

\end{tabular}
\end{center}
\caption{Transition}
\label{transition}
\end{figure}

The new architecture evolved through much trial-and-error experimentation.
Much of the axiomatization for the metashell 
has been taken from 
some core IFF meta-ontologies in the old architecture.
In fact,
loosely speaking,
we have effectively raised 
the previous four-tiered (ur-top-upper-lower) architecture of the old core 
to become the new metashell (Figure~\ref{transition}).
But,
this happened somewhat accidently with no a prior design.
We have been guided only by the requirements.
And apparently these requirements have intrinsically remained the same,
but only shifted upwards in the metalevel hierarchy.
For the {\small\ttfamily IFF-IFF} namespace,
we have used the recent version of the IFF Ur (meta) Ontology (July 2004)\footnote{\,located at \url{http://suo.ieee.org/IFF/metastack/UR.pdf}}
(but simplified,
since we have moved introduction of the predicate and relation concepts 
down to the type level).
For the type kernel namespace,
we have used
a version of the IFF Top Core (meta) Ontology 
called the IFF Basic KIF (meta) Ontology (January 2002)\footnote{\,located at \url{http://suo.ieee.org/IFF/metalevel/top/ontology/core/version20020102.pdf}}.
For the diagram and finite-limit namespaces at the type level,
we have used
the version of the IFF Top Core (meta) Ontology in adjunctive style axiomatization (June 2004)\footnote{\,located at \url{http://suo.ieee.org/IFF/metastack/TCO.pdf}}.
For the meta kernel namespace, and the diagram and finite-limit namespaces at the meta level,
we have used
a version of the IFF Upper Core (meta) Ontology (January 2002)\footnote{\,located at \url{http://suo.ieee.org/IFF/metalevel/upper/ontology/core/version20020102.pdf}}.
For the exponent namespace at the meta level,
we have used
the stub for the IFF Upper Core (meta) Ontology in adjunctive style axiomatization (July 2004)\footnote{\,located at \url{http://suo.ieee.org/IFF/metastack/UCO.pdf}}.

\subsubsection{Syntax}\label{subsubsec-syntax}

\begin{table}
\begin{center}
\begin{tabular}{c}
\\
\fbox{\begin{tabular}{l}
{\scriptsize \begin{tabular}{rrl}
& {\small\sffamily Math} & {\small\sffamily IFF}\rule[3pt]{0pt}{10pt} \\ 
\multicolumn{3}{l}{{\sffamily names}\rule[3pt]{0pt}{10pt}} \\
& $a$ & {\scriptsize\ttfamily o.i:a} \makebox[16pt][l]{\begin{tabular}{r} inner context $i$ \\ 
outer context $o$ \end{tabular}} \\
\multicolumn{3}{l}{{\sffamily atoms and terms}\rule[-5pt]{0pt}{12pt}}             \\
set element          & $x \in X$ or $X(x)$       & {\scriptsize\ttfamily (X x)}   \\
predicate member     & $x \in b$ or $b(x)$       & {\scriptsize\ttfamily (b x)}   \\
relation member      & $(x,y) \in r$ or $r(x,y)$ & {\scriptsize\ttfamily (r x y)} \\
function application & $f(x)$                    & {\scriptsize\ttfamily (f x)}   \\ 
\multicolumn{3}{l}{{\sffamily equations}\rule[-4pt]{0pt}{16pt}}                   \\
         & $\sigma = \tau$ or ${=}(\sigma,\tau)$ & {\scriptsize\ttfamily (= s t)} \\ 
\multicolumn{3}{l}{{\sffamily connectives}\rule[-4pt]{0pt}{16pt}}                             \\
conjunction & $\phi \wedge \psi$ or ${\wedge}(\phi,\psi)$ & {\scriptsize\ttfamily (and P Q)}  \\ 
disjunction & $\phi \vee \psi$ or ${\vee}(\phi,\psi)$     & {\scriptsize\ttfamily (or P Q)}   \\ 
implication & $\phi \Rightarrow \psi$ or ${\Rightarrow}(\phi,\psi)$ 
                                                      & {\scriptsize\ttfamily (implies P Q)}  \\ 
equivalence & $\phi \Leftrightarrow \psi$ or ${\Leftrightarrow}(\phi,\psi)$ 
                                                          & {\scriptsize\ttfamily (iff P Q)}  \\ 
negation    & $\neg \phi$ or ${\neg}(\phi)$               & {\scriptsize\ttfamily (not P)}    \\ 
\multicolumn{3}{l}{{\sffamily quantifiers}\rule[-4pt]{0pt}{16pt}}                             \\
universal   & $\forall_{x_0 \in X_0, x_1 \in X_1} \phi$\hspace{10pt} &                        \\
            & or $\forall_{(x_0 \in X_0, x_1 \in X_1)} (\phi)$ 
            & {\scriptsize\ttfamily (forall ((X0 x0) (X1 x1)) P)}                             \\
existential & $\exists_{x_0 \in X_0, x_1 \in X_1} \phi$\hspace{10pt} & \rule[-1pt]{0pt}{10pt} \\
            & or $\exists_{(x_0 \in X_0, x_1 \in X_1)} (\phi)$ 
            & {\scriptsize\ttfamily (exists ((X0 x0) (X1 x1)) P)} \rule[-5pt]{0pt}{10pt}      \\
\end{tabular}}
\end{tabular}}
\end{tabular}
\end{center}
\caption{Syntax Tutorial}
\label{syntax:tutorial}
\end{table}

The LISt Processing (LISP) programming language,
which is based on the lambda calculus,
is the second oldest (1958) programming language
--- only the FORTRAN language is older.
LISP became the favored programming language for artificial intelligence research.
All program code is written as parenthesized lists. 
The Knowledge Interchange Format (KIF),
which has a LISP-like format,
was created to serve as a syntax for first-order logic. 
The IFF logical notation (Table~\ref{syntax:tutorial}), 
which is a vastly simplified and modified version of KIF, 
also has a LISP-like format.
The IFF grammar\footnote{located at \url{http://suo.ieee.org/IFF/grammar.pdf}}
is written in Extended Backus Naur Form (EBNF).
The IFF language contains both logical code and comments.
Both nested namespaces and metalevels are specified by prefixes.
Table~\ref{example:code} illustrates some IFF code examples.
On the left side is code taken from the metashell.
The first example 
taken from the {\small\ttfamily IFF-IFF} namespace
states that two sets are equal precisely when they have the same elements.
The second and third examples, 
taken from the function and predicate subnamespaces of {\small\ttfamily IFF-TYPE},
define the belonging and membership relations,
respectively, 
as discussed in Section~\ref{sec-foundations} on foundations.
On the right side is code 
taken from the core ontology ({\small\ttfamily IFF-SET}) in the natural part.
This illustrates the chain of code at various metalevels that is necessary 
in order to declare a level $n$ function
$f \in \mathsf{mor}(\mathsf{Set}_{n})$.
Note that the metashell code is first order,
but the natural part code is (almost) atomic,
consisting of either declarations, equations or relational expressions (and one negation).

\begin{table}
\begin{center}
\begin{tabular}{@{\hspace{0pt}}l@{\hspace{0pt}}l}
\begin{tabular}[t]{r@{\hspace{0pt}}l}
{\tiny\ttfamily \begin{tabular}[t]{l}
iff:
\end{tabular}}
&
\fbox{\tiny\ttfamily \begin{tabular}[t]{l}
(forall ((set ?X) (set ?Y))                 \\
\hspace{10pt}
    (iff (= ?X ?Y)                          \\
\hspace{26pt}
         (and (forall ((?X ?x)) (?Y ?x))    \\ 
\hspace{41pt}
              (forall ((?Y ?y)) (?X ?y)))))
\end{tabular}}
\\ \\
{\tiny\ttfamily \begin{tabular}[t]{l}
type.ftn:
\end{tabular}}
&
\fbox{\tiny\ttfamily \begin{tabular}[t]{l}
(iff:set belonging)                                            \\
(forall ((belonging ?xy))                                      \\
\hspace{10pt}
    (type.dgm.pr.mor:function-pair ?xy))                       \\
(forall ((type.ftn:function ?x)                                \\ 
\hspace{26pt}
         (type.ftn:function ?y))                               \\
\hspace{10pt}
    (iff (belonging [?x ?y])                                   \\
\hspace{26pt}
         (exists ((2-cell ?a))                                 \\
\hspace{39pt}
             (and (= ?x (source ?a))                           \\
\hspace{55pt}
                  (= ?y (target ?a)))))) 
\end{tabular}}
\\ \\
{\tiny\ttfamily \begin{tabular}[t]{l}
type.pred:
\end{tabular}}
&
\fbox{\tiny\ttfamily \begin{tabular}[t]{l}
(iff:set membership) \\
(forall ((membership ?xp)) \\
\hspace{10pt}
    (exists ((type.ftn:function ?x) (predicate ?p)) \\
\hspace{23pt}
        (= ?xp [?x ?p]))) \\
(forall ((type.ftn:function ?x) (predicate ?p)) \\
\hspace{10pt}
    (iff (membership [?x ?p]) \\
\hspace{26pt}
         (type.ftn:belonging [?x (function ?p)])))
\end{tabular}}
\end{tabular}
&
\begin{tabular}[t]{r@{\hspace{0pt}}l}
{\tiny\ttfamily \begin{tabular}[t]{l}
\#n.set:
\end{tabular}}
&
\fbox{\tiny\ttfamily \begin{tabular}[t]{l}
((\#n+1).set:set set)                \\
((\#n+2).set:subset set (\#n+1).set:set) \\
(not (set set))
\end{tabular}}
\\ \\
{\tiny\ttfamily \begin{tabular}[t]{l}
\#n.ftn:
\end{tabular}}
&
\fbox{\tiny\ttfamily \begin{tabular}[t]{l}
((\#n+1).set:set function)                       \\
((\#n+2).set:subset function (\#n+1).ftn:function)   \\ \\
((\#n+1).ftn:function source)                    \\
(= ((\#n+1).ftn:source source) function) \\
(= ((\#n+1).ftn:target source) \#n.set:set)      \\
((\#n+2).ftn:restriction source (\#n+1).ftn:source)  \\ \\
((\#n+1).ftn:function target)                    \\
(= ((\#n+1).ftn:source target) function) \\
(= ((\#n+1).ftn:target target) \#n.set:set)      \\
((\#n+2).ftn:restriction target (\#n+1).ftn:target)
\end{tabular}}
\\ \\
{\tiny\ttfamily \begin{tabular}[t]{l}
abc:
\end{tabular}}
&
\fbox{\tiny\ttfamily \begin{tabular}[t]{l}
(\#n.set:set A)          \\
(\#n.set:set B)          \\
(\#n.ftn:function f)     \\
(= (\#n.ftn:source f) A) \\
(= (\#n.ftn:target f) B)
\end{tabular}}
\end{tabular}
\\ \\ \multicolumn{1}{c}{\sffamily metashell} & \multicolumn{1}{c}{\sffamily natural part}
\end{tabular}
\end{center}
\caption{Example Code}
\label{example:code}
\end{table}

\subsubsection{Metalanguage}\label{subsubsec-metalanguage}

In logic and linguistics, 
a metalanguage is a language used 
to make statements about other languages (object languages)\footnote{Merriam-Webster,Wikipedia}. 
Here,
we use the term metalanguage 
to mean a language used to formally express the meaning of another language.
An ordered metalanguage is a sequence of metalanguages
where each is used as a metalanguage for the next (previous) one.
Each level represents a lesser (greater) degree of abstraction. 
A nested, or hierarchical, metalanguage is an ordered metalanguage 
where each level is included in (includes) the next one. 
Each IFF metalevel 
$1,2,\ldots n,\ldots,{\mbox{\small$\mathtt{meta}$}},{\mbox{\small$\mathtt{type}$}},{\mbox{\small$\mathtt{iff}$}}$
services the levels below by providing a metalanguage 
used to declare and axiomatize those levels. 
Each of these metalanguages is used to organize and axiomatize the metalanguages below it. 
Hence,
the IFF is a nested, or hierarchical, metalanguage (see \cite{kent:04} for further discussion)
\begin{center}
{\footnotesize $\begin{array}{l}
\mathtt{IFF} = \left\langle
\mathtt{IFF}_{\mathtt{iff}} \subset 
\mathtt{IFF}_{\mathtt{type}} \subset 
\mathtt{IFF}_{\mathtt{meta}} \subset 
\cdots \subset 
\mathtt{IFF}_{n{+}1} \subset 
\mathtt{IFF}_{n} \subset 
\cdots
\mathtt{IFF}_{2} \subset 
\mathtt{IFF}_{1} 
\right\rangle.
\end{array}$}
\end{center}
The {\small\ttfamily iff} level metalanguage {\small $\mathtt{IFF}_{\mathtt{iff}}$}, 
which is used to talk about (axiomatize) type things, 
is also used to talk about (axiomatize) meta things and level $n$ things;
that is,
{\small $\mathtt{IFF}_{\mathtt{iff}} \subset \mathtt{IFF}_{\mathtt{type}}$}\footnote{The metalanguage {\footnotesize $\mathtt{IFF}_{\mathtt{iff}}$} is very simple,
consisting of only the logical symbols in the IFF grammar,
plus the five terms
$\{\mbox{`{\scriptsize\ttfamily thing}'},
\mbox{`{\scriptsize\ttfamily set}'},
\mbox{`{\scriptsize\ttfamily function}'},
\mbox{`{\scriptsize\ttfamily source}'},
\mbox{`{\scriptsize\ttfamily target}'}\}$
defined in the {\footnotesize\ttfamily IFF-IFF} namespace.
The metalanguage {\footnotesize $\mathtt{IFF}_{\mathtt{type}}$} is more complicated,
consisting of the logical symbols, 
these five {\footnotesize\ttfamily iff} terms,
and around 450 other terms defined in the {\footnotesize\ttfamily IFF-TYPE} namespace
that are needed to define a finitely-complete category 
of abstract sets and functions.
The entire IFF metalanguage,
which changes as new terminology is added,
consists of thousands of terms.}.
The {\small\ttfamily type} level metalanguage {\small $\mathtt{IFF}_{\mathtt{type}}$}, 
which is used to talk about (axiomatize) meta things, 
is also used to talk about (axiomatize) level $n$ things;
that is,
{\small $\mathtt{IFF}_{\mathtt{type}} \subset \mathtt{IFF}_{\mathtt{meta}}$}.
The level $n{+}1$ metalanguage {\small $\mathtt{IFF}_{n{+}1}$}, 
which is used to talk about (axiomatize) level $n$ things, 
is also used to talk about (axiomatize) level $m$ things for all $m<n$;
that is,
{\small $\mathtt{IFF}_{n{+}1} \subset \mathtt{IFF}_{n}$}.
In the other direction,
level $n$ metalevel namespaces and meta-ontologies 
use level $m$ terminology and functionality for any metalevel $m$, where $n < m$;
the {\small\ttfamily meta} namespace 
uses the {\small\ttfamily type} and {\small\ttfamily iff} level terminology and functionality; 
and the {\small\ttfamily type} namespace 
uses the {\small\ttfamily iff} level terminology and functionality.

\section{Foundations}\label{sec-foundations}

\begin{flushleft}
{\small \begin{tabular}{p{4.8in}}
\emph{``All of the substance of mathematics can be fully expressed in categories.''} \\
$\sim$ F.W.~Lawvere \rule{0pt}{10pt}
\end{tabular}}
\end{flushleft}
The metashell, 
core component,
and structural component of the natural part
provide 
pointwise (non-traditional) set-theoretic,
pointless set-theoretic\footnote{Words can have different meanings for different groups in different contexts.
The term `point' (or the equivalent term `element') could have a different meaning 
in the set-theoretic context 
than it would 
in the category-theoretic context.
The category-theoretic context might regard a morphism to be a (generalized) element,
and consider the case of an element of source 1 to be a special element.
The set-theoretic context might restrict its notion of element to only these special elements.
This is the intent in the jest above.},
and category-theoretic foundations
for the IFF, respectively.
An alternate foundations can be chosen by disconnecting the metashell from the natural part 
(severing suitable metastack linkage) and discarding it.
The core and structural components of the natural part 
correspond to the distinction between the foundations and organization of mathematics 
as discussed in the presentation \emph{Kreisel and Lawvere} by J.P. Marquis\footnote{see \url{http://www.math.mcgill.ca/rags/seminar/Marquis_KreiselLawvere.pdf}}. 
This component architecture of the IFF, 
which is forced by the principles of conceptual warrant and categorical design, 
agrees with Marquis's claim that ``We can have both worlds!''.
The foundational approach of the IFF,
using core and structural components,
possibly addresses some concerns of S. Fefermann \cite{feferman:77}
concerning the logical and psychological priority of the notions of collection and operation 
(over categorical notions). 

\subsection{Misconceptions}\label{subsec-misconceptions}

Although the IFF is principally oriented towards the roles of applications and support,
since the intended support is towards a standard for category theory,
the IFF has inevitably found itself involved in the philosophical role of foundations.
In this section, 
our approach to the IFF philosophical role uses the two misconceptions 
about category theory discussed by F.W.~Lawvere 
in three messages to the CAT list entitled \emph{Why are we concerned?}\footnote{copy located at 
\url{http://categorytheorynews.blogspot.com/}}.
\begin{description}
\item [Misconception 1:]
``Category theory is the `insubstantial part' of mathematics and 
it heralds an era when precise axioms are no longer needed.''
This misconception was discussed in Lawvere's second message.
\item [Misconception 2:] 
``There are `size problems' if one tries to do category theory 
in a way harmonious with the standard practice of professional set theorists.''
This misconception was discussed in Lawvere's third message.
\end{description}
We handle these in reverse order.
The discussion on the Cantorian expansion in subsection~\ref{subsec-cantorian:expansion}
addresses the second misconception.
The discussion on inclusion and membership in subsection~\ref{subsec-inclusion:membership}
addresses the first misconception.
As a side comment,
according to Lawvere in his second message,
``Contrary to Fregean rigidity, 
in mathematics we never use `properties' that are defined on the universe of `everything'.
There is the `universe of discourse' principle which is very important: 
for example, any given group, (or any given topological space, etc.)
acts as a universe of discourse.''
The IFF syntax addresses this issue.
It requires the use of \emph{restricted quantification} in logical expression.
For example,
the following IFF code axiomatizes the inverse element for a group: 
{\scriptsize\begin{verbatim}
(forall ((group ?G))
    (forall ((?G ?a)) 
        (and (= ((multiplication ?G) [?a ((inverse ?G) ?a)]) (unit ?G))
             (= ((multiplication ?G) [((inverse ?G) ?a) ?a]) (unit ?G)))))
\end{verbatim}}

\subsection{Cantorian Expansion}\label{subsec-cantorian:expansion}

The book \emph{Sets for Mathematics} 
by Lawvere and Rosebrugh \cite{lawvere:rosebrugh:03}
has a discussion and a closely reasoned argument (proof) for Cantor's theorem.
In this subsection 
we extend that argument resulting in a structure we call the Cantorian expansion.
The Cantorian expansion provides the metastack,
which is the spine of the IFF architecture.

\subsubsection{Cantor's Theorem}\label{subsubsec-cantors:theorem}

Let $Y$ be any set. 
An element $y \in Y$ is a fixed point of an endofunction $\tau : Y \rightarrow Y$ 
when $\tau(y) = y$. 
A set $Y$ has the fixed point property 
when every endofunction on $Y$ has at least one fixed point.
Suppose there is a set $X$ and a function $\varphi : X{\times}X \rightarrow Y$ 
whose curry $\hat{\varphi} : X \rightarrow Y^X$, 
where $\hat{\varphi}(a) = \varphi(a,\mbox{-})$ for all $a \in X$, is surjective; 
that is, 
such that for every function $f : X \rightarrow Y$ 
there is at least one element $a \in X$ such that $f = \hat{\varphi}(a) = \varphi(a,\mbox{-})$. 
Then $Y$ has the fixed point property.

\begin{flushleft}
{Theorem:}
If a set $Y$ has at least one endofunction $\tau : Y \rightarrow Y$ with no fixed points, 
then for every set $X$ there is no surjection $X \rightarrow Y^X$.
In particular, 
letting $Y = 2$,
since negation $\neg : 2 \rightarrow 2$ has no fixed points,
$X < 2^X = \mbox{\large$\wp$}X$ for any set $X$.
That is,
every set $X$ is strictly smaller than its powerset $2^X = \mbox{\large$\wp$}X$.
\end{flushleft}

\noindent
As a corollary,
there cannot exist a ``universal set'' $U$ for which every set $X$ is a subset $X \subseteq U$:
if so, 
then the inclusion $X \rightarrow U$ is an injection;
hence, 
the exponent map $2^U \rightarrow 2^X$ is a surjection; 
then we can define $X = 2^U$ to get a contradiction.
As a further corollary,
the collection $\mathsf{set}$ of all sets is not a set:
if $\mathsf{set}$ were a set, 
then $U = \mbox{\footnotesize $\bigcup$}\, \mathsf{set}$ would be a ``universal set''. 
There are two essential ingredients for Cantor's theorem and the above corollaries:
existence of 
the powerset function $\mbox{\large$\wp$}$ and 
the (unbounded) union function {\footnotesize $\bigcup$}.

\subsubsection{Cantorian Expansion}\label{subsubsec-cantorian:expansion}

The sets here are called ``small'' sets.
The collection of small sets,
like the set of natural numbers $\aleph$,
is either defined naturally, by convention or
logically/mathematically\footnote{The set of natural numbers,
which occurs in nature,
was used in antiquity (convention)
and axiomatized in modern times (logic/math).}.
The last corollary states that there are sets that are not small.
Change the notation,
letting $\mathsf{set}_1$ denote the collection of small sets,
and $\mathsf{set}_2$ denote the collection of sets either small or not 
(call them ``large'' sets).
So that
$\mathsf{set}_1 \subseteq \mathsf{set}_2$,
$\mathsf{set}_1 \in \mathsf{set}_2$, but
$\mathsf{set}_1 \not\in \mathsf{set}_1$.
This argument is relative and can be repeated
by next using the large sets as the ``new'' notion of small sets.
Hence,
starting from the small sets $\mathsf{set}_1$,
the collection of all sets unfolds into a chain 
\begin{center}
{\small \begin{tabular}{r@{\hspace{5pt}$=$\hspace{5pt}}l}
$\mathsf{set}$ 
& 
$\underbrace{\left\langle
\mathsf{set}_1 \subset \mathsf{set}_2 \subset \cdots \subset \mathsf{set}_n \subset \cdots  
\right\rangle}$
\\
\multicolumn{1}{c}{} & \multicolumn{1}{c}{Cantorian expansion}
\end{tabular}}
\end{center}
(of Cantorian featureless abstract sets)
called the \emph{Cantorian expansion}.

\subsubsection{Smallness}\label{subsubsec-smallness}

The Cantorian expansion is used as the foundation of the IFF. 
It is the minimal assumption required to obey Cantor's theorem. 
In the IFF the collection of small sets is defined by convention\footnote{To reiterate,
the Cantorian expansion is by convention.
First, 
we assume by convention that a set of ``small'' sets exists.
Call this $\mathsf{set}_1$.
We also assume there are power and unbounded union operators defined on $\mathsf{set}_1$.
By Cantor's theorem $\mathsf{set}_1 \not\in \mathsf{set}_1$.
Hence,
there are sets that are not just small.
Call such sets, including $\mathsf{set}_1$, ``properly large'' sets.
Call a set, that is either small or properly large, a ``large'' set. 
Second, 
we assume by convention that a set of ``large'' sets exists.
Call this $\mathsf{set}_2$.
We also assume there are power and unbounded union operators defined on $\mathsf{set}_2$.
Note that
$\mathsf{set}_1 \in \mathsf{set}_2$,
$\mathsf{set}_1 \subset \mathsf{set}_2$ and
$\mathsf{set}_1 \not\in \mathsf{set}_1$.
By Cantor's theorem $\mathsf{set}_2 \not\in \mathsf{set}_2$.
Hence,
there are sets that are not just large.
Call such sets, including $\mathsf{set}_1$, ``properly very large'' sets.
Call a set, that is either small, large or properly very large, a ``very large'' set. 
Third, 
we assume by convention that a set of ``very large'' sets exists.
Call this $\mathsf{set}_3$.
We also assume there are power and unbounded union operators defined on $\mathsf{set}_3$.
Note that
$\mathsf{set}_2 \in \mathsf{set}_3$,
$\mathsf{set}_2 \subset \mathsf{set}_3$ and
$\mathsf{set}_2 \not\in \mathsf{set}_2$.
By Cantor's theorem $\mathsf{set}_3 \not\in \mathsf{set}_3$.
Hence,
there are sets that are not just very large.
And we continue in this fashion,
thereby generating the Cantorian expansion.}.
F.W.~Lawvere has suggested at the end of the third email message above that 
the collection of small sets could be defined logically/mathematically.
In particular,
that the small sets could be constructed from the large sets as 
the solution to a certain set isomorphism (equation) expressed using the real numbers. 
Since the IFF follows only a minimal approach to foundations, 
it should be possible to further constrain the IFF axiomatization 
to respect Lawvere's definition of smallness.

\subsubsection{Unions and Universes}\label{subsubsec-unions:universes}

Let $n$ be any metalevel
with $\mathsf{set}_{n}$ the collection of all level $n$ sets\footnote{Following 
the Cantorian expansion,
we assume that
$\mathsf{set}_{n}$ is closed under subset order
($X \in \mathsf{set}_{n}, Y \subseteq X$ implies
$Y \in \mathsf{set}_{n}$),
any level $n$ set is a level $n{+}1$ set
($\mathsf{set}_{n} \subseteq \mathsf{set}_{n{+}1}$),
$\mathsf{set}_{n}$ is itself a level $n{+}1$ set
($\mathsf{set}_{n} \in \mathsf{set}_{n{+}1}$)
and
$\mathsf{set}_{n}$ is not a level $n$ set
($\mathsf{set}_{n} \not\in \mathsf{set}_{n}$).}.
For any level $n$ set $X \in \mathsf{set}_{n}$, 
the bounded union operation\footnote{In the current version of the IFF,
although we have bounded unions at every metalevel,
except for Cantor 
we have not used either unbounded unions or universes.} 
$\bigcup_{n}^{X} 
: \mbox{\large${\wp}$}\mbox{\large${\wp}$}X \rightarrow \mbox{\large${\wp}$}X$
is defined as
$\bigcup_{n}^{X}(Z) 
= \{ x \in X \mid \exists_Y (x \in Y \in Z) \}$
for 
any family of subsets 
$Z \in \mbox{\large${\wp}$}\mbox{\large${\wp}$}X$.
Define
$\mbox{\large$\tilde{\wp}$}(\mathsf{set}_{n})
= \mathsf{set}_{n} {\cap}\, \mbox{\large${\wp}$}\mathsf{set}_{n}
\in \mathsf{set}_{n{+}1}$
to be the collection of level $n$ sets of level $n$ sets.
Then,
$\mbox{\large${\wp}$}\mbox{\large${\wp}$}X
\subseteq 
\mbox{\large$\tilde{\wp}$}(\mathsf{set}_{n})$. 
The unbounded union operation\footnote{In the IFF, 
this union would be specified within and local to a particular IFF metalevel.}
$\bigsqcup_{n} 
: \mbox{\normalsize$\tilde{\wp}$}(\mathsf{set}_{n}) \rightarrow \mathsf{set}_{n}$
is defined as 
$\bigsqcup_{n}(Z) 
= \{ x \mid \exists_Y (x \in Y \in Z) \}$\footnote{Following Mac Lane in the foundations section of \emph{Categories for the Working Mathematician} \cite{maclane:71}.}
for any family
$Z \in \mbox{\large$\tilde{\wp}$}(\mathsf{set}_{n})$.
Since
$\bigcup_{n}^{X}(Z) 
= \bigsqcup_{n}\!Z$
for any $Z \in \mbox{\large${\wp}$}\mbox{\large${\wp}$}X$,
the bounded union is the restriction of the unbounded union.
A level $n$ universe is a level $n{+}1$ set $U \in \mathsf{set}_{\mathsf{n{+}1}}$
that has the properties:
$\mathsf{set}_{n} \subseteq U$
``every level $n$ set is an element of the universe''
and
$\mathsf{set}_{n} \subseteq \mbox{\large${\wp}$}U$
``every level $n$ set is a subset of the universe''.
Then,
$\mbox{\large$\tilde{\wp}$}(\mathsf{set}_{n})
\subseteq \mbox{\large${\wp}$}\mbox{\large${\wp}$}U$.
Define the specific level $n$ universe
$\mathsf{univ}_{n}
= \bigsqcup_{\mathsf{n{+}1}} \mathsf{set}_{n}
= \{ x \mid \exists_Y (x \in Y \in \mathsf{set}_{n}) \}
\in \mathsf{set}_{\mathsf{n{+}1}}$.
This defines the special bounded union function 
$\bigcup_{n{+}1}^{\mathsf{univ}_{n}} : \mbox{\normalsize${\wp}$}\mbox{\normalsize${\wp}$}\mathsf{univ}_{n} \rightarrow \mbox{\normalsize${\wp}$}\mathsf{univ}_{n}$.
Then,
the unbounded union of level $n$ sets is the restriction of this bounded union. 

\subsubsection{Grothendieck Universes}\label{subsubsec-grothendieck:universes}

\hspace{-5pt}The IFF has much in common with Grothendieck universes.
A Grothendieck universe $\mathcal{U}$ is meant to provide a set 
in which all of mathematics can be performed\footnote{(by way of Wikipedia)
Bourbaki, N., 
\emph{Univers}, 
appendix to Exposé I of Artin, M., Grothendieck, A., Verdier, J. L., eds., 
\emph{Th\'{e}orie des Topos et Cohomologie \'{E}tale des Sch\'{e}mas (SGA 4)}, 
second edition, Springer-Verlag, Heidelberg, 1972.}.
The IFF provides a framework in which all of mathematics can be axiomatized.
Grothendieck universes model universes of sets.
However,
IFF universes contain 
non-set objects such as
functions, predicates, relations, 
vectors, numbers, 
ships, stars, pelicans and bacteria.
This means that Grothendieck universes are more like the toposes 
$\langle \mathsf{Set}_{n}, 1 \leq n < \infty \rangle$
than the IFF universes
$\langle \mathsf{univ}_{n}, 1 \leq n < \infty \rangle$.
Indeed,
the main intuition is that for any set $X$, 
there is a Grothendieck universe $\mathcal{U}$ with $X \in \mathcal{U}$. 
Similarly,
for any IFF set $X$, 
there is a whole number $1 \leq n < \infty$ 
with $X \in \mathsf{set}_{n} = \mathsf{obj}(\mathsf{Set}_{n})$.
More precisely,
a Grothendieck universe $\mathcal{U}$ is a set 
which is closed under membership and 
contains doubletons, powers and indexed unions.
These axioms imply that
a Grothendieck universe $\mathcal{U}$ is closed under the subset order, 
and contains functions, isomorphs, singletons, 
indexed coproducts (disjoint unions), 
indexed products and indexed intersections.
Analogs between Grothendieck universes and the IFF are listed in 
Table~\ref{grothendieck:iff:analogs}.

\begin{table}
\begin{center}
\fbox{\footnotesize \begin{tabular}{|p{180pt}@{\hspace{10pt}}p{180pt}|} 
\multicolumn{1}{l}{\sffamily\normalsize Grothendieck universe} 
& \multicolumn{1}{l}{\sffamily\normalsize IFF} \rule[-6pt]{0pt}{12pt} \\ \hline
\makebox[0pt][r]{{\sffamily Axioms:}\hspace{12pt}}\hspace{-3.5pt}
If $X \in \mathcal{U}$ and $x \in X$, then $x \in \mathcal{U}$.
& If $X \in \mathsf{set}_{n}$ and $x \in X$, then $x \in \mathsf{univ}_{n}$. \rule[0pt]{0pt}{12pt} \\
If $x,y \in \mathcal{U}$, then $\{x,y\} \in \mathcal{U}$.
& doubleton function
$\{\mbox{-},\mbox{-}\}_X : X^2 \rightarrow \mbox{\normalsize${\wp}$}(X)$ \\
If $X \in \mathcal{U}$, then $\mbox{\normalsize${\wp}$}(X) \in \mathcal{U}$.
& power set function
$\mbox{\normalsize${\wp}$} : \mathsf{set}_{n} \rightarrow \mathsf{set}_{n}$ \\
If $I \in \mathcal{U}$ and $X_\alpha \in \mathcal{U}$ for each $\alpha \in I$, 
then $\bigcup_{\alpha \in I} X_\alpha \in \mathcal{U}$.
& bounded union
$\bigcup_{n}^{X} : \mbox{\normalsize${\wp}$}\mbox{\normalsize${\wp}$}X \rightarrow \mbox{\normalsize${\wp}$}X$,
and unbounded union
$\bigsqcup_{n} : \mbox{\normalsize$\tilde{\wp}$}(\mathsf{set}_{n}) \rightarrow \mathsf{set}_{n}$
\rule[-6pt]{0pt}{12pt} \\ \hline\hline
\makebox[0pt][r]{{\sffamily Theorems:}\hspace{12pt}}\hspace{-3.5pt}
If $X \in \mathcal{U}$ and $Y \subseteq X$, then $Y \in \mathcal{U}$. 
& If $X \in \mathsf{set}_{n}$ and $Y \subseteq X$, then $Y \in \mathsf{set}_{n}$. \rule[0pt]{0pt}{12pt} \\
If $X,Y \in \mathcal{U}$, then $(f : X \rightarrow Y) \in \mathcal{U}$.
& If $X,Y \in \mathsf{set}_{n}$, then $(f : X \rightarrow Y) \in \mathsf{ftn}_{n}$. \\
If $X \in \mathcal{U}$ and $Y \cong X$, then $Y \in \mathcal{U}$.
& If $X \in \mathsf{set}_{n}$ and $Y \cong X$, then $Y \in \mathsf{set}_{n}$. \\
If $x \in \mathcal{U}$, then $\{x\} \in \mathcal{U}$.
& singleton function $\{\mbox{-}\}_X : X \rightarrow \mbox{\normalsize${\wp}$}(X)$ \\
If $I \in \mathcal{U}$ and $X_\alpha \in \mathcal{U}$ for each $\alpha \in I$, then 
$\coprod_{\alpha \in I} X_\alpha,  \prod_{\alpha \in I} X_\alpha,
 \bigcap_{\alpha \in I} X_\alpha \in \mathcal{U}$. \rule[-6pt]{0pt}{12pt} 
& The category $\mathsf{set}_{n}$ is small (co)complete.
The preorder $\mbox{\normalsize${\wp}$}(X)$ is a Boolean algebra. \\ \hline
\end{tabular}}
\end{center}
\caption{Grothendieck-IFF Analogs}
\label{grothendieck:iff:analogs}
\end{table}

\subsection{Inclusion and Membership}\label{subsec-inclusion:membership}

\begin{table}
\begin{center}
\begin{tabular}[t]{c@{\hspace{10pt}}|@{\hspace{20pt}}c@{\hspace{10pt}}}
\begin{tabular}{c|c|c|}
\multicolumn{1}{c}{} & \multicolumn{2}{c}{{\footnotesize (type)}} 
\rule[-6pt]{0pt}{10pt}
\\
\multicolumn{1}{c}{} & \multicolumn{2}{c}{$\mathsf{set}$}
\\
\multicolumn{1}{c}{} &
\multicolumn{2}{c}{\setlength{\unitlength}{0.35pt}
\begin{picture}(0,0)(5,-10)
\put(0,0){\framebox(0,0){}}
\put(0,0){\circle*{5}}
\end{picture}}
\rule[-6pt]{0pt}{10pt}
\\ \cline{2-3}
&
$\mathsf{ftn}$
&
$\mathsf{pred}$
\rule[0pt]{0pt}{10pt}
\\
\begin{tabular}{c}{\footnotesize 1-dim}\\{\footnotesize (unary)}\end{tabular}
&
\setlength{\unitlength}{0.35pt}
\begin{picture}(80,44)(6,-10)
\put(10,-18){\framebox(60,40){}}
\put(17,-10){\makebox(50,20){$\Downarrow$}}
\end{picture}
&
\setlength{\unitlength}{0.35pt}
\begin{picture}(80,44)(6,-10)
\put(10,-18){\framebox(60,40){}}
\put(17,-10){\makebox(50,20){$\rule{0.3pt}{4pt}\;\!{\scriptstyle \wedge}$}}
\end{picture}
\rule[-8pt]{0pt}{10pt}
\\ \cline{2-3}
&
$\mathsf{spn}$
&
$\mathsf{rel}$
\rule[0pt]{0pt}{10pt}
\\
\begin{tabular}{c}{\footnotesize 2-dim}\\{\footnotesize (binary)}\end{tabular}
&
\setlength{\unitlength}{0.35pt}
\begin{picture}(80,44)(4,-10)
\put(-10,-28){\framebox(100,60){}}
\qbezier(0,5)(40,45)(80,5)
\put(80,5){\vector(1,-1){0}}
\put(17,-10){\makebox(50,20){$\Downarrow$}}
\qbezier(0,-5)(40,-45)(80,-5)
\put(80,-5){\vector(1,1){0}}
\end{picture}
&
\setlength{\unitlength}{0.35pt}
\begin{picture}(80,44)(2,-10)
\put(-10,-28){\framebox(100,60){}}
\qbezier(0,5)(40,45)(80,5)
\put(80,5){\vector(1,-1){0}}
\put(17,-10){\makebox(50,20){$\rule{0.3pt}{4pt}\;\!{\scriptstyle \wedge}$}}
\qbezier(0,-5)(40,-45)(80,-5)
\put(80,-5){\vector(1,1){0}}
\end{picture}
\rule[-10pt]{0pt}{10pt}
\\ \cline{2-3}
\multicolumn{1}{c}{} & \multicolumn{1}{c}{{\footnotesize (element)}} & \multicolumn{1}{c}{{\footnotesize (part)}}
\rule[0pt]{0pt}{12pt}
\end{tabular}
{\hspace{40pt}}
&
\begin{tabular}{c}
\setlength{\unitlength}{0.55pt}
\begin{picture}(100,120)(0,0)
\put(-25,90){\makebox(50,20){$\mathsf{ftn}$}}
\put(-25,-10){\makebox(50,20){$\mathsf{spn}$}}
\put(75,90){\makebox(50,20){$\mathsf{pred}$}}
\put(75,-10){\makebox(50,20){$\mathsf{rel}$}}
\put(25,11){\makebox(50,20){\scriptsize{$\mathsf{rel}$}}}
\put(25,-4){\makebox(50,20){\scriptsize{$\in$}}}
\put(25,-19){\makebox(50,20){\scriptsize{$\mathsf{spn}$}}}
\put(25,111){\makebox(50,20){\scriptsize{$\mathsf{pred}$}}}
\put(25,96){\makebox(50,20){\scriptsize{$\in$}}}
\put(25,81){\makebox(50,20){\scriptsize{$\mathsf{ftn}$}}}
\put(-40,40){\makebox(50,20){\scriptsize{$\mathsf{ftn}$}}}
\put(33,40){\makebox(50,20){\scriptsize{$\mathsf{emb}$}}}
\put(90,40){\makebox(50,20){\scriptsize{$\mathsf{pred}$}}}
\put(-40,105){\makebox(50,20){\scriptsize{$\sqsubseteq$}}}
\put(90,105){\makebox(50,20){\scriptsize{$\subseteq$}}}
\put(-40,-25){\makebox(50,20){\scriptsize{$\sqsubseteq$}}}
\put(90,-25){\makebox(50,20){\scriptsize{$\subseteq$}}}
\put(25,150){\makebox(50,20){\scriptsize{$\mathsf{ext}(\in)$}}}
\put(25,-70){\makebox(50,20){\scriptsize{$\mathsf{ext}(\in)$}}}
\put(-27,137){\makebox(50,20){\scriptsize{$\mathsf{prf}$}}}
\put(-90,-40){\makebox(50,20){\scriptsize{$\mathsf{prf}$}}}
\put(20,15){\vector(1,0){60}}
\put(20,0){\line(1,0){60}}
\put(80,-15){\vector(-1,0){60}}
\put(80,-19){\oval(8,8)[r]}
\put(20,115){\vector(1,0){60}}
\put(20,100){\line(1,0){60}}
\put(80,85){\vector(-1,0){60}}
\put(80,81){\oval(8,8)[r]}
\put(0,25){\vector(0,1){50}}
\put(-4,25){\oval(8,8)[b]}
\put(100,25){\vector(0,1){50}}
\put(96,25){\oval(8,8)[b]}
\put(-16,115){\oval(30,30)[t]}
\put(-16,115){\oval(30,30)[bl]}
\put(-1,115){\line(0,-1){6}}
\put(116,115){\oval(30,30)[t]}
\put(116,115){\oval(30,30)[br]}
\put(101,115){\line(0,-1){6}}
\put(-16,-15){\oval(30,30)[b]}
\put(-16,-15){\oval(30,30)[tl]}
\put(-1,-15){\line(0,1){6}}
\put(116,-15){\oval(30,30)[b]}
\put(116,-15){\oval(30,30)[tr]}
\put(101,-15){\line(0,1){6}}
\put(33,150){\vector(-2,-3){20}}
\put(67,150){\vector(2,-3){20}}
\put(33,-50){\vector(-2,3){20}}
\put(67,-50){\vector(2,3){20}}
\put(6,120){\vector(1,-4){0}}
\qbezier(23,155)(0,145)(6,120)
\put(-13,80){\vector(1,2){0}}
\qbezier(20,-58)(-110,-60)(-13,80)
\put(50,150){\vector(-1,2){0}}
\qbezier(50,-45)(95,50)(50,150)
\put(46,-45){\oval(8,8)[b]}
\end{picture}
\end{tabular}
\\ & \\ \hline
\begin{tabular}[t]{cc}
& \\
\rule[4pt]{0pt}{12pt}
\begin{tabular}[t]{c}
\small\ttfamily{ftn}
\\
\setlength{\unitlength}{0.5pt}
\begin{picture}(60,100)(0,-10)
\put(-25,30){\makebox(50,20){\footnotesize{$X$}}}
\put(5,62){\makebox(50,20){\scriptsize{$x$}}}
\put(5,-2){\makebox(50,20){\scriptsize{$y$}}}
\put(45,30){\makebox(50,20){\scriptsize{$p$}}}
\put(60,68){\vector(0,-1){56}}
\put(51,74){\vector(-3,-2){40}}
\put(51,6){\vector(-3,2){40}}
\end{picture}
\\
\footnotesize\sffamily{slice category}
\end{tabular}
&
\begin{tabular}[t]{c}
\small\ttfamily{pred}
\\
\setlength{\unitlength}{0.5pt}
\begin{picture}(60,100)(0,-10)
\put(-25,30){\makebox(50,20){\footnotesize{$X$}}}
\put(5,62){\makebox(50,20){\scriptsize{$p$}}}
\put(5,-2){\makebox(50,20){\scriptsize{$q$}}}
\put(64,58){\oval(8,8)[t]}
\put(60,58){\vector(0,-1){44}}
\qbezier(51,74)(58,75)(55,68)
\put(51,74){\vector(-3,-2){40}}
\qbezier(51,6)(55,-1)(47,0)
\put(51,6){\vector(-3,2){40}}
\end{picture}
\\
\footnotesize\sffamily{subobject preorder}
\end{tabular}
\\ \\
\begin{tabular}[t]{c}
\ttfamily{spn}
\\
\setlength{\unitlength}{0.5pt}
\begin{picture}(120,100)(0,-10)
\put(-25,30){\makebox(50,20){\footnotesize{$X_0$}}}
\put(95,30){\makebox(50,20){\footnotesize{$X_1$}}}
\put(3,62){\makebox(50,20){\scriptsize{$x_0$}}}
\put(3,-2){\makebox(50,20){\scriptsize{$y_0$}}}
\put(45,30){\makebox(50,20){\scriptsize{$p$}}}
\put(72,62){\makebox(50,20){\scriptsize{$x_1$}}}
\put(72,-2){\makebox(50,20){\scriptsize{$y_1$}}}
\put(60,68){\vector(0,-1){56}}
\put(51,74){\vector(-3,-2){40}}
\put(51,6){\vector(-3,2){40}}
\put(69,74){\vector(3,-2){40}}
\put(69,6){\vector(3,2){40}}
\end{picture}
\\
\footnotesize\sffamily{span bicategory}
\end{tabular}
&
\begin{tabular}[t]{c}
\ttfamily{rel}
\\
\setlength{\unitlength}{0.5pt}
\begin{picture}(60,100)(0,-10)
\put(-25,30){\makebox(50,20){\footnotesize{$X_0{\times}X_1$}}}
\put(5,62){\makebox(50,20){\scriptsize{$r$}}}
\put(5,-2){\makebox(50,20){\scriptsize{$s$}}}
\put(64,58){\oval(8,8)[t]}
\put(60,58){\vector(0,-1){44}}
\qbezier(51,74)(58,75)(55,68)
\put(51,74){\vector(-3,-2){35}}
\qbezier(51,6)(55,-1)(47,0)
\put(51,6){\vector(-3,2){35}}
\end{picture}
\\
\footnotesize\sffamily{relation 1.5-category}
\end{tabular}
\end{tabular}
& 
\begin{tabular}[t]{l}
$\begin{array}[t]{|r@{\hspace{3pt}\scriptstyle:\hspace{3pt}}c@{\hspace{3pt}\scriptstyle\rightarrow\hspace{3pt}}c|r|}
\hline
  \multicolumn{3}{|c|}{\mbox{\footnotesize\sffamily Concept}} 
& \multicolumn{1}{|c|}{\mbox{\footnotesize\sffamily IFF Term}} \\ \hline\hline
\scriptstyle\mathsf{pred} & \scriptstyle\mathsf{ftn} & \scriptstyle\mathsf{pred} 
& \mbox{\scriptsize\ttfamily predicate} \\
\scriptstyle\mathsf{ftn} & \scriptstyle\mathsf{pred} & \scriptstyle\mathsf{ftn} 
& \mbox{\scriptsize\ttfamily function} \\
\scriptstyle\mathsf{rel} & \scriptstyle\mathsf{spn} & \scriptstyle\mathsf{rel} 
& \mbox{\scriptsize\ttfamily relation} \\
\scriptstyle\mathsf{spn} & \scriptstyle\mathsf{rel} & \scriptstyle\mathsf{spn} 
& \mbox{\scriptsize\ttfamily span} \\ \hline
\scriptstyle\mathsf{ftn} & \scriptstyle\mathsf{spn} & \scriptstyle\mathsf{ftn} 
& \mbox{\scriptsize\ttfamily function} \\
\scriptstyle\mathsf{pred} & \scriptstyle\mathsf{rel} & \scriptstyle\mathsf{pred} 
& \mbox{\scriptsize\ttfamily predicate} \\
\hline
\end{array}$
\\
{\scriptsize $\begin{array}[t]{ll}
\rule[4pt]{0pt}{10pt}
\mbox{$\mathsf{ftn}$-$\mathsf{pred}$ reflection:}              &                               \\
f \stackrel{\eta}{\Rightarrow} \mathsf{ftn}(\mathsf{pred}(f)), & \forall f \in \mathsf{ftn}    \\
p \cong \mathsf{pred}(\mathsf{ftn}(p)),                        & \forall p \in \mathsf{pred}   \\
                                                               &                               \\
\mbox{$\mathsf{spn}$-$\mathsf{rel}$ reflection:}               &                               \\
s \stackrel{\eta}{\Rightarrow} \mathsf{spn}(\mathsf{rel}(s)),  & \forall s \in \mathsf{spn}    \\
r \cong \mathsf{rel}(\mathsf{spn}(r)),                         & \forall r \in \mathsf{rel}    \\
                                                               &                               \\
\mbox{$\mathsf{pred}$-$\mathsf{sub}$ equivalence:}             &                               \\
X\!Y = \mathsf{sub}(\mathsf{pred}(X\!Y)),                      & \forall X\!Y \in \mathsf{sub} \\
p \cong \mathsf{pred}(\mathsf{sub}(p)),                        & \forall p \in \mathsf{pred}
\end{array}$}
\end{tabular}
\\ & \\ 
\end{tabular}
\end{center}
\caption{IFF Subobject Architecture}
\label{subobject:architecture}
\end{table}

According to Lawvere,
the first misconception ``is connected with taking seriously the jest `sets without elements'.'' 
We first give the usual definition of basic concepts,
such as inclusion and membership, 
in the category-theoretic context.
This follows the usual mathematical practice of defining these relations 
for subsets of a given universe of discourse represented by an arbitrary category.
Then we discuss where these are most generally represented in the IFF.

\subsubsection{Basic Definitions}\label{subsubsec-basic:definitions}

Let $\mathcal{C}$ be any category
and let $X$ be any $\mathcal{C}$-object.
For any $\mathcal{C}$-morphism $x$, 
$x$ is an element of $X$, 
symbolized $x \in X$, 
when $X$ is the target of $x$.
For any $X$-element $b \in X$, 
$b$ is a part of $X$\footnote{In the IFF, 
$b$ is called a predicate with genus $X$.},
symbolized $b : X$,
when $b$ is a monomorphism. 
For any two $X$-elements $x,y \in X$, 
$x$ belongs to $y$,
symbolized $x \Rightarrow y$,
when there exists a proof $\mathcal{C}$-morphism $p$ 
such that $x = p \cdot y$\footnote{Here, composition is written in diagrammatic order.}.
When $y$ is an $X$-part,
the proof $p$ of that belonging is unique.
For any two $X$-parts $a,b : X$, 
$a$ is included in $b$,
symbolized $a \subseteq b$, 
when $a \Rightarrow b$. 
For any $X$-element $x \in X$
and any $X$-part $b : X$, 
$x$ is a member of $b$,
symbolized $x \in b$, 
when $x \Rightarrow b$.
As Lawvere points out,
the usual relationship holds between inclusion and membership;
that is,
inclusion is equivalent to universal implication of membership:
$a \subseteq b \;\;\mbox{iff}\;\; \forall_{x \in X} (x \in a \;\;\mbox{implies}\;\;x \in b)$
for any two X-parts $a,b : X$.

\subsubsection{The Type Namespace}\label{subsubsec-type:namespace}


The {\small\ttfamily IFF-TYPE} namespace,
a small-sized namespace at the top of the IFF architecture,
is at the heart of the IFF metashell.
Since the metashell unfolds into the rest of the IFF,
we could regard the {\small\ttfamily IFF-TYPE} namespace,
and its kernel in particular, 
as the heart of the entire IFF.
The {\small\ttfamily IFF-TYPE} namespace is in the middle of the metashell,
just below the {\small\ttfamily IFF-IFF} namespace.
But the {\small\ttfamily IFF-IFF} namespace has little actual content (axiomatization),
since it only specifies set and function kind of things.
The {\small\ttfamily IFF-TYPE} namespace is where the axiomatization really begins.
It has three nested subnamespaces:
the type kernel namespace and namespaces for type diagrams and type limits.
The {\small\ttfamily IFF-TYPE} namespace defines the finitely-complete category 
$\mathsf{Set}_{\mathsf{type}}$ 
of abstract sets and functions
enriched with factorization and subobjects.
In a standard fashion, 
the finitely-complete category $\mathsf{Set}_{\mathsf{type}}$ defines
the bicategory 
$\mathsf{Span}_{\mathsf{type}} = \mathsf{spn}(\mathsf{Set}_{\mathsf{type}})$ 
of type sets and type spans,
the ordered category 
$\mathsf{Rel}_{\mathsf{type}} = \mathsf{rel}(\mathsf{Set}_{\mathsf{type}})$ 
of type sets and type (binary) relations,
and their categorical connections.
In addition to canonical functionality for finite limits,
the {\small\ttfamily IFF-TYPE} namespace, 
specifically its kernel subnamespace,
defines (Table~\ref{subobject:architecture}) 
four related subobject structures,
the belonging relation for functions and spans, and 
the inclusion relation for predicates and relations.
Also defined are membership relations between functions and predicates,
and between spans and relations.
Associated with these two membership relations are maps that return the proof of membership.

\subsubsection{Generalized Composition}\label{subsubsec-generalized:composition}

Composition and generalized elements are two essential ingredients in the IFF development.
The IFF uses category theory,
both in its applications and as a foundation for its architecture.
Since composition is fundamental in category theory,
it is also fundamental in the use and development of the IFF architecture.
In particular,
composition for generalized elements
is crucially important.
Not only do we use the usual composition operation 
$\circ^n : \mathsf{ftn}_n{\times}_{\mathsf{set}_n}\mathsf{ftn}_n \rightarrow \mathsf{ftn}_n$
for ordinary point pairs (composable pairs of level $n$ functions)
$(f : 1 \rightarrow \mathsf{ftn}_n, g : 1 \rightarrow \mathsf{ftn}_n)$
returning a level $n$ function
$f \circ^n g : 1 \rightarrow \mathsf{ftn}_n$,
but we also use the parametric composition operation 
$\hat{\circ}^n$
for generalized point pairs (composable level $n{+}1$ spans)
$(f : \mathsf{ftn}_n \leftarrow X \rightarrow \mathsf{ftn}_n : g)$
for some level $n$ (vertex) set $X$
returning a level $n{+}1$ function
$f \hat{\circ}^n g = (f,g)\, \circ^{n{+}1} \circ^n : X \rightarrow \mathsf{ftn}_n$.
The use of composition for generalized elements 
is vital to ensure that the IFF natural part has an atomic expression.
Such generalized composition has its most general definition in the meta kernel namespace.
This definition uses 
the type level belonging relation for spans and relations
along with its associated proof operation.

\section{Future Work}\label{sec-future:work}

All scientific communities, 
indeed all communities of discourse (disciplines), 
create their own conceptual structures 
with accompanying terminology and meaning (ontologies).
Many communities are now working to standardize their ontologies.
There is also a search for a unifying framework for these endeavers. 
It has been suggested \cite{dampney:johnson:01}
that category theory can serve this role 
--- category theory can serve as a meta-ontology, an ontology of ontologies. 
This was the goal of the SUO IFF project as initially envisioned in the year 2000.

\subsection{The IFF}\label{subsec-future:work:iff}

Following the two guiding principles of conceptual warrant and categorical design,
through much experimentation over the last six years
the IFF architecture has been developed in essentially a bottom-up fashion.
For example\footnote{Actually, the very first IFF ontology developed was the IFF Category Theory meta-ontology {\scriptsize\ttfamily IFF-CAT} \cite{kent:01} located at 
\url{http://suo.ieee.org/IFF/metalevel/upper/ontology/category-theory/version20020102.pdf}.},
following the principle of conceptual warrant 
various ontologies representing first order logic 
were axiomatized in a first order expression,
and then higher level meta-structure was axiomatized which,
following the principle of categorical design,
would move the axiomatization closer to an atomic expression.
During this experimental process\footnote{The motto is that ``the IFF is an experiment in foundations''.} much logical code was developed for the IFF.

\subsubsection{Logical Coding}\label{subsubsec-logical:coding}

\begin{figure}
\begin{center}
\begin{tabular}{c@{\hspace{50pt}}c}
\begin{tabular}{c}
\setlength{\unitlength}{0.9pt}
\begin{picture}(100,50)(0,0)
\put(-10,40){\makebox(0,0)[r]{\sffamily\scriptsize{metashell}}}
\put(-10,19){\makebox(0,0)[r]{\sffamily\scriptsize{natural}}}
\put(-10,8){\makebox(0,0)[r]{\sffamily\scriptsize{part}}}
\put(18,31){\makebox(0,0){\sffamily\tiny{core}}}
\put(12,40){\makebox(0,0){\sffamily\tiny{$1.$}}}
\put(20,40){\makebox(0,0){\sffamily\tiny{$2.$}}}
\put(28,40){\makebox(0,0){\sffamily\tiny{$3.$}}}
\put(20,19){\makebox(0,0){\sffamily\tiny{$4.$}}}
\put(46,19){\makebox(0,0){\sffamily\tiny{$5.$}}}
\put(80,9){\makebox(0,0){\sffamily\tiny{$6.$}}}
\put(91,9){\makebox(0,0){\sffamily\tiny{$7.$}}}
\put(18,11){\makebox(0,0){\sffamily\tiny{core}}}
\put(45,11){\makebox(0,0){\sffamily\tiny{struc}}}
\put(30,-1){\makebox(0,0){\sffamily\scriptsize{pure}}}
\put(85,-1){\makebox(0,0){\sffamily\scriptsize{applied}}}
\put(0,50){\vector(0,-1){20}}
\put(0,25){\vector(0,-1){25}}
\put(10,35){\vector(1,0){20}}
\put(30,31){\oval(8,8)[r]}
\put(10,27){\line(1,0){22}}
\put(10,15){\vector(1,0){24}}
\put(10,21){\oval(8,12)[l]}
\put(34,15){\vector(1,0){21}}
\put(55,11){\oval(8,8)[tr]}
\put(63,11){\oval(8,12)[bl]}
\put(10,5){\vector(1,0){45}}
\put(63,5){\vector(1,0){42}}
\end{picture}
\end{tabular}
&
{\tiny\ttfamily \begin{tabular}{lr@{-}l}
1. & IFF & IFF \\
2. &     & TYPE \\
3. &     & META \\
4. &     & SET($n$) \\
5. &     & CAT($n$), \ldots \\
6. &     & INS \\
7. &     & FOL, \ldots \\
   & \multicolumn{1}{r}{} & \multicolumn{1}{l}{\ldots}
\end{tabular}}
\end{tabular}
\end{center}
\caption{IFF Coding Process Flow}
\label{process:flow}
\end{figure}
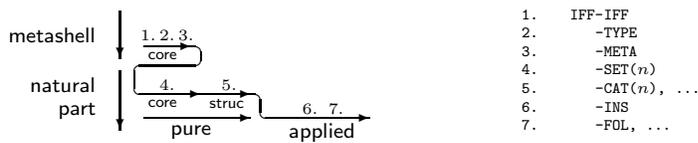

Recently the final IFF coding was initiated.
Great reuse will be made of the experimental code.
The coding process flow will follow the architecture in a top-down inside-out fashion as illustrated in Figure~\ref{process:flow}.
The axiomatization (1,2) of 
the {\small\ttfamily IFF-IFF} and {\small\ttfamily IFF-TYPE} namespaces is complete.
Coding of the {\small\ttfamily IFF-IFF} namespace was trivial,
since it only involved five terms: 
`{\small\ttfamily thing}', `{\small\ttfamily set}', `{\small\ttfamily function}', `{\small\ttfamily source}' and `{\small\ttfamily target}'.
However because of its importance,
the {\small\ttfamily IFF-TYPE} namespace that has approximately 450 terms was carefully coded.
The next phase of work is (3) the {\small\ttfamily IFF-META} namespace.
Ignoring exponents,
the meta kernel namespace is conceptually the specialization of the type kernel namespace.
However,
in the {\small\ttfamily IFF-META} namespace 
the axiomatic expressions are flatter and more atomic.
Also,
the predicate/relation hold expressions,
which were defined in the {\small\ttfamily IFF-TYPE} namespace, 
are in great use.
The concurrent development processes 
(illustrated in Figure~\ref{current:development:state})
need to be finished in the meta kernel.
After the {\small\ttfamily IFF-META} namespace is axiomatized,
work will proceed on (4) the generic (level $n$) set ontology [{\small\ttfamily IFF-SET}],
and then (5) the structural component of the IFF 
[{\small\ttfamily IFF-CAT}, {\small\ttfamily IFF-2CAT}, {\small\ttfamily IFF-DCAT}, etc.].
After the pure aspect is sufficiently coded.
work will continue on (6) the ontology axiomatization for institution theory 
[{\small\ttfamily IFF-INS}].
We will also integrate with this (7) previous axiomatizations for logic
[{\small\ttfamily IFF-OBJ}, {\small\ttfamily IFF-FOL}, {\small\ttfamily IFF-IF}]. 

\subsubsection{Implementation}\label{subsubsec-implementation}

Implementation of the IFF will aid the working category theorist.
The first step of implementation is finished.
This involved writing the IFF grammar. 
The IFF grammar 
(written in EBNF)\footnote{see the IFF syntax document at 
\url{http://suo.ieee.org/IFF/grammar.pdf}} 
works for all of the IFF (both pure and applied aspects, and the metashell).
Logical coding of the IFF is based upon this grammar.
Also based upon the IFF grammar,
we will be able to realize the implementation of a tool suite for the IFF 
(parsers, viewers, editors, reasoners, etc.). 
With such an implementation, there could be various application strategies. 
One strategy might be to represent in the IFF both the syntax and semantics of various other frameworks (common logic, OWL, etc.) in order to absorb any and all of their coded information. 
But there could be other possible strategies.

\subsection{The Community}\label{subsec-community}

Much of the previous activities concerning the IFF involved 
the application of category theory to knowledge engineering, the semantic web,
and the representation and semantic integration of ontologies.
However,
we now want to reverse this approach.
We want to apply knowledge engineering to support category theory.
A critical element for future work on the IFF involves 
the development of a standard ontology for the category theory community.

\subsubsection{Standards}\label{subsubsec-standards}

A standard is something established by authority, custom, or general consent 
to be used as a model, example or rule.
There are many reasons for standards.
Standards allow interoperability between cooperating groups in technology, business and science.
Standards are intended to be documented, known descriptions of how something works 
so that every group who adheres to the standard can interoperate. 
There is no one true standard.
In fact, the evolution of standards is a sign of healthy innovation.
The plurality of standards-issuing organizations 
means that some standards do not necessarily have the support of all communities.
The standards of large communities can be created
to replace the various incompatible standards of smaller communities.

\subsubsection{Kinds}\label{subsubsec-kinds}

There are several kinds of standards.
Two distinctions for standards are 
open versus proprietary and de facto versus de jure.
An open standard is documented for all,
and developed and maintained by peers and in the public arena.
A proprietary standard is developed and maintained by/for one particular organisation
(e.g. Microsoft's Windows, Adobe's PDF).
A de facto standard is the property of consortia that represent a wide range of interests.
Its status is conferred by use in the marketplace.
(e.g. IETF's HTTP protocol, W3C's HTML format, OMG's CORBA)
A de jure standard is developed by standards bodies 
established under national or international laws.
It is a well-documented convention, 
agreed to by participants in a formal standards forum,
created through a formal process, and 
based on the work of a cooperative group or committee of experts.
(e.g. the meter of the French Academy of Sciences\footnote{one/ten-millionth of the distance from the equator to the north pole}
and the Geneva Conference on Weights and Measures\footnote{the distance light travels in a vacuum in 1/299,792,458 seconds},
ANSI's C programming language,
ISO's JPEG).
Also,
an ad hoc standard is more widely used than their originator intended
(e.g. JVC's VHS, Compuserve's GIF).
Various standards project of the Institute of Electrical and Electronics Engineers (IEEE),
such as the SUO project, 
are good examples of the standardization process.
The IEEE is heavily involved in developing and maintaining standards for current technologies.
Table~\ref{standards:process} summarizes the IEEE standization principles and process.

\paragraph{Proposal}\label{para-proposal}

\emph{The category theory community will form a working group 
(under the auspices of some organization or consortium) 
for the purpose of developing a standard ontology for category theory.}

\begin{table}
\begin{center}
\begin{tabular}{cc}
{\scriptsize \fbox{\begin{tabular}[t]{l}
\multicolumn{1}{l}{\bfseries Principles} \rule[-6pt]{0pt}{15pt} \\
$\bullet$ \emph{due process} \\
$\bullet$ \emph{openness}    \\
$\bullet$ \emph{consensus}   \\
$\bullet$ \emph{balance}     \\
$\bullet$ \emph{right of appeal}
\end{tabular}}}
&
{\scriptsize \fbox{\begin{tabular}[t]{l}
\multicolumn{1}{l}{\bfseries Process} \rule[-6pt]{0pt}{15pt} \\
0. idea \\
1. project approval by standards body \\
2. draft development by working group \\
  \hspace{16pt} a. form WG with chair and technical editor \\
  \hspace{16pt} b. establish goals, deadlines and schedule \\
  \hspace{16pt} c. draft document                          \\
  \hspace{16pt} d. reviewed by technical editor            \\
3. sponsor ballot                     \\
4. standards board approval process   \\
5. publish standards                  \\
6. periodically reaffirm, revise or withdraw standard; goto 1.
\end{tabular}}}
\end{tabular}
\end{center}
\caption{IEEE Standards}
\label{standards:process}
\end{table}


\bibliographystyle{plain}
\bibliography{kent}

\end{document}